\documentclass[aps,prd,reprint,twocolumn,superscriptaddress,longbibliography,nofootinbib,floatfix,showpacs]{revtex4-2}

\usepackage{epsfig}
\usepackage{amsmath,amssymb,amsfonts}
\usepackage{hyperref}
\usepackage{mathrsfs}
\usepackage{bbm}
\usepackage{slashed}
\usepackage{graphicx}
\usepackage{verbatim}

\usepackage{bm}

\usepackage[usenames]{color}

\definecolor{darkgreen}{rgb}{0.2,0.6,0}

\newcommand{\be}{\begin{equation}}
\newcommand{\ee}{\end{equation}}
\newcommand{\bw}{\begin{widetext}}
\newcommand{\ew}{\end{widetext}}
\newcommand{\bi}{\begin{itemize}}
\newcommand{\ei}{\end{itemize}}
\newcommand{\ud}{\mathrm{d}}

\newcommand{\LCperp}{{\scriptscriptstyle \perp}}

\usepackage[T1]{fontenc} \usepackage[latin1]{inputenc}

\begin{document}

\title{Quantum radiation reaction: Analytical approximations and obtaining the spectrum from moments}

\author{Greger Torgrimsson}
\email{greger.torgrimsson@umu.se}
\affiliation{Department of Physics, Ume{\aa} University, SE-901 87 Ume{\aa}, Sweden}

\begin{abstract}

We derive analytical $\chi\ll1$ approximations for spin-dependent quantum radiation reaction for locally constant and locally monochromatic fields. We show how to factor out fast spin oscillations and obtain the degree of polarization in the plane orthogonal to the magnetic field from the Frobenius norm of the Mueller matrix. We show that spin effects lead to a transseries in $\chi$, with powers $\chi^k$, logarithms $(\ln\chi)^k$ and oscillating terms, $\cos(\dots/\chi)$ and $\sin(\dots/\chi)$. 
In our approach we can obtain each moment, $\langle(kP)^m\rangle$, of the lightfront longitudinal momentum independently of the other moments and without considering the spectrum. 
We show how to obtain a low-energy expansion of the spectrum from the moments by treating $m$ as a continuous, complex parameter and performing an inverse Mellin transform. We also show how to obtain the spectrum, without making a low-energy approximation, from a handful of moments using the principle of maximum entropy.  

\end{abstract}
\maketitle

\section{Introduction}

A common approach to quantum radiation reaction (RR) is to use kinetic equations~\cite{Shen1972,Sokolov:2010am,Neitz:2013qba,Neitz:2014hla,Ridgers2017,NielPRE2018,Elkina:2010up}, where the old rates for constant-crossed fields~\cite{Nikishov:1964zza,Ritus:1985vta} are used as ingredients. See~\cite{Gonoskov:2021hwf} for a review, and~\cite{DiPiazza:2011tq,Fedotov:2022ely} for other aspects of RR and strong-field QED. In recent years people have realized that spin and polarization effects can be important~\cite{DelSorbo:2017fod,DelSorbo2,Seipt:2019ddd,Chen:2019vly,Li:2018fcz} and a great deal of work has been done to generalize the previous methods to include spin and polarization, e.g. in particle-in-cell (PIC) codes~\cite{Li:2018fcz,Li:2019oxr,Li:2020bwo,Tang:2021azl,Li:2022wqn,Qian:2023tqc}. Stokes vectors have emerged as a useful tool for describing general spin and polarization states. In~\cite{Dinu:2018efz,Dinu:2019pau,Torgrimsson:2020gws} we calculated\footnote{We are using the Furry picture, so a $\mathcal{O}(\alpha)$ term is still a nontrivial/nonperturbative function of the background field $eF^{\mu\nu}$.} $\mathcal{O}(\alpha)$ strong-field-QED Mueller matrices and showed how one can obtain the dominant contribution of higher-orders-in-$\alpha$ processes from products of these matrices. In~\cite{Torgrimsson:2021wcj,Torgrimsson:2021zob,Torgrimsson:2022ndq,Torgrimsson:2023rqo} we showed how to resum the resulting $\alpha$ expansion, resulting in some integro-differential equations for RR. Mueller matrices have now also been incorporated into kinetic equations~\cite{Seipt:2023bcw}.

This incoherent-product approximation is valid if $a_0$ is large enough and/or the pulse sufficiently long. Here\footnote{We use units with $m_e=c=\hbar=1$ and absorb $eF^{\mu\nu}\to F^{\mu\nu}$.} $a_0=E/\omega$, $E$ is the maximum field strength and $1/\omega$ is a characteristic length scale of the field. In this paper we will neglect pair production, which is exponentially suppressed for small $\chi$, where $\chi=\sqrt{-(F^{\mu\nu}p_\nu)^2}$ is the quantum nonlinearity parameter.  

Recently, \cite{Blackburn:2023dey} calculated analytical approximations at next-to-leading order (NLO) in $\chi\ll1$ in the locally-constant-field (LCF) regime $a_0\gg1$. 
In this paper we will calculate analytical approximations at NLO for the resummed Mueller matrix of different moments and for both the LCF regime and for more general regimes, where the incoherent product of $\mathcal{O}(\alpha)$ Mueller matrices still give the dominant contribution. If $a_0\sim1$ then this is the case if the pulse is sufficiently long. If the field is almost monochromatic then one can treat it using a locally-monochromatic-field (LMF) approximation~\cite{Dinu:2013hsd,Heinzl:2020ynb,Torgrimsson:2020gws}, see also~\cite{Seipt:2010ya}. Such ideas have also been included in some numerical codes~\cite{Blackburn:2023mlo}. This is easier to do for a circularly polarized field. Another reason for considering circular polarization is that spin effects do not average out~\cite{Torgrimsson:2021zob}, which they would tend to do for linear polarization.    
One can also avoid spin effects averaging out by considering various asymmetric oscillations~\cite{Seipt:2019ddd,Chen:2019vly} or unipolar fields, which can be motivated e.g. by beam-beam collisions~\cite{Song:2021wou,Gao:2024vlf}.

We choose coordinates so that the plane wave background field is described by a wave vector $k_\mu=\omega(1,0,0,1)$ and a polarization vector $a_\mu(\phi)=\{0,a_1,a_2,0\}$, which is in general a pulsed function of lightfront time $\phi=kx=\omega(t+z)$. 
In a plane wave the lightfront longitudinal momentum $kP$ plays a key role, where $P_\mu$ is the momentum of the electron. We can study the dynamics of $kP$ without considering the transverse components of the momentum $P_\LCperp$. We are in particular interested in the moments $\langle(kP)^m\rangle$. In our approach we obtain this from a Mueller matrix $\tilde{\bf M}(m)$. 
In~\cite{Torgrimsson:2021zob,Torgrimsson:2022ndq,Torgrimsson:2023rqo} we derived the following integro-differential equation for the moments,
\be\label{MnEq}
\frac{\partial}{\partial\sigma}\tilde{\bf M}(b_0)=
-\int_0^1\!\ud q\bigg\{{\bf M}^L\!\cdot\!\tilde{\bf M}(b_0)+{\bf M}^C\!\cdot\!\tilde{\bf M}\left([1-q]b_0\right)\bigg\} \;,
\ee
where $\sigma$ is a lightfront time variable, $b_0=kp$ is the initial longitudinal momentum, $q=kl/b_0$ and $l_\mu$ is the momentum of the emitted or loop photon, and we have ``initial'' conditions at $\sigma\to+\infty$,
\be\label{initialMn}
\tilde{\bf M}(\sigma\to+\infty,m,b_0)=b_0^m{\bf 1}_{4\times4} \;.
\ee
${\bf M}^C$ is the Mueller matrix coming from nonlinear Compton scattering, and ${\bf M}^L$ comes from the electron-mass-operator loop. The reason ${\bf M}^L$ contributes at the same order as ${\bf M}^C$ is because ${\bf M}^L$ comes from the cross term between the $\mathcal{O}(\alpha^0)$ and $\mathcal{O}(\alpha)$ terms in the amplitude~\cite{Torgrimsson:2020gws}. It follows from unitarity that some of the elements of ${\bf M}^L$ are equal to the corresponding elements in ${\bf M}^C$ but with opposite sign. However, ${\bf M}^L$ also contains terms that are not only numerically different from ${\bf M}^C$ but also have a significantly different mathematical structure, as we will show below.      
We only need to consider the initial and final spin states at the very end of the calculation, where we obtain the moments by projecting onto the initial (${\bf N}_0$) and final (${\bf N}_f$) Stokes vectors for the electron spin,
\be\label{N0MNf}
\langle(kP)^m\rangle=\frac{1}{2}{\bf N}_0\cdot\tilde{\bf M}\cdot{\bf N}_f \;.
\ee

Note that, in contrast to the usual kinetic equations, with~\eqref{MnEq} we treat each moment separately. There is no hierarchy relating one moment to other moments and we do not need to compute the spectrum. 
So, if we want to calculate e.g. the second moment then we only have to consider the second moment. 
Another thing to note is that~\eqref{MnEq} is integrated backwards in (lightfront) time. Thus, while we showed in~\cite{Torgrimsson:2023rqo} that~\eqref{MnEq} is consistent with the standard kinetic approach, \eqref{MnEq} provides a genuinely different approach. Given that we can obtain the moments independently, a relevant question is to what extent we can reproduce the spectrum from knowledge of the moments. In this paper we study two different ways of doing so, by 1) treating $m$ as a complex Mellin variable or 2) using the principle of maximum entropy.       

This paper is organized as follows. In Sec.~\ref{LCF of moments} we calculate the first couple of terms in the $\chi\ll1$ expansion of the moments $\tilde{\bf M}(m)$ in the LCF regime, and explain two methods for calculating the exact $\tilde{\bf M}$ numerically. In Sec.~\ref{Mellin transform} we show how to obtain the $\chi\ll1$ expansion of the spectrum by treating $m$ as a complex, Mellin integration variable. In Sec.~\ref{Maximum entropy} we show how to use the principle of maximum entropy to obtain the spectrum from $\tilde{\bf M}(m)$ with only a handful of moments $m$. In Sec.~\ref{Beyond LCF} we calculate the leading quantum correction beyond the LCF regime, i.e. for a sufficiently long field but without assuming $a_0\gg1$. In Sec.~\ref{Spin perpendicular to magnetic field} we show how to factor out fast spin oscillations for a linearly polarized field in the LCF regime and obtain the degree of polarization from the Frobenius norm of the Mueller matrix.

\section{LCF approximation of moments}\label{LCF of moments}

We first consider the LCF regime, $a_0\gg1$.
We use the same idea as in Eq.~(77) in~\cite{Torgrimsson:2023rqo}, i.e. we first factor out the overall $b_0^m$ scaling as
\be
\tilde{\bf M}(\sigma,m,b_0)=:b_0^m{\bf W}(\sigma,m,b_0) \;.
\ee
From~\eqref{initialMn} we have
\be
{\bf W}(\sigma\to+\infty,m,b_0)={\bm 1} \;,
\ee
i.e. the same ``initial'' conditions for all moments. Different moments ${\bf W}(m)$ and ${\bf W}(m')$ are now typically on the same order of magnitude (unless some of the components vanish identically to some order, which we will see examples of below).
The equation of motion now depends explicitly on $m$,
\be\label{Weq}
\begin{split}
\frac{\partial}{\partial\sigma}{\bf W}(b_0)=&-\int_0^1\!\ud q\bigg\{{\bf M}^L\cdot{\bf W}(b_0)\\
&+(1-q)^m{\bf M}^C\cdot{\bf W}([1-q]b_0)\bigg\} \;.
\end{split}
\ee 
At this point we have not yet assumed the LCF regime. In LCF, the Mueller matrices ${\bf M}^{L,C}$ are given by Eqs.~(53) and~(54) in~\cite{Torgrimsson:2023rqo}, which are expressed in terms of a locally-constant value of the quantum nonlinearity parameter, 
\be\label{chiF}
\chi(\sigma)=|{\bf a}'(\sigma)|b_0=\chi_0 F(\sigma) \;,
\ee
where $\chi_0=a_0b_0$ is the maximum value of $\chi$, and $F$ is a dimensionless function, e.g. $F(x)=e^{-x^2}$. We consider for simplicity a linearly polarized field, so that the magnetic field is always either parallel or antiparallel to a constant, unit vector $\hat{\bf B}_0$,
\be
\hat{\bf B}(\sigma)=\epsilon(\sigma)\hat{\bf B}_0 \;,
\ee
where $\epsilon(\sigma)=\pm1$ depending on the sign of $a'(\sigma)/|a'(\sigma)|$. The 4D Stokes vectors live in a vector space that is spanned by $\hat{\bf B}_0=-{\bf e}_2$, the electric field direction $\hat{\bf E}_0={\bf e}_1$, the propagation direction $\hat{\bf k}={\bf e}_3$, and a vector that describes an unpolarized electron ${\bf e}_0$. For a linearly polarized field, the ${\bf e}_0$-$\hat{\bf B}_0$ components decouple from the $\hat{\bf E}_0$-$\hat{\bf k}$ components. In this section we will only consider the ${\bf e}_0$-$\hat{\bf B}_0$ components. We consider the $\hat{\bf E}_0$-$\hat{\bf k}$ components in Sec.~\ref{Spin perpendicular to magnetic field}.

In the constant-field case~\cite{Torgrimsson:2023rqo}, we defined a ``time'' or pulse-length parameter $u$, such that the longitudinal component of the solution to the LL equation is given by $kP_{LL}/b_0=1/(1+u)$, i.e. $u=(2/3)\chi\alpha a_0\Delta\sigma$. Note that $u$ is proportional to $\chi$ or $b_0$, but we want to consider $u=\mathcal{O}(1)$ even when $\chi\ll1$, because otherwise we would not see the nontrivial dependence on $u$. Now when we consider non-constant fields, we cannot use the same $u$ variable, but we still want to capture the nontrivial dependencies. To do so we define
\be
\rho:=\frac{2}{3}\alpha b_0 a_0^2 \;,
\ee 
which we consider to be $\mathcal{O}(1)$, which is the case if $a_0$ is sufficiently large to compensate for $\alpha\ll1$ and $b_0\ll1$. Our results below are still valid if $\rho$ is smaller, but then they simply reduce to $\mathcal{O}(\alpha)$.
Then we expand the solution as a power series in $\chi_0$,
\be\label{Wsumw}
{\bf W}(\sigma,\chi_0,m)=\sum_{k=0}^\infty{\bf w}_{m,k}(\sigma,\rho)\chi_0^k \;,
\ee
but where the coefficients as are still nontrivial functions of $\rho$. We will show in Sec.~\ref{Spin perpendicular to magnetic field} that the $\hat{\bf E}_0$-$\hat{\bf k}$ components of ${\bf W}$ also contain terms with $\ln\chi_0$ and $\cos(.../\chi_0)$ or $\sin(.../\chi_0)$.
From ${\bf M}^{L,C}$ we have an overall factor of $\alpha/b_0=(3/2)(\rho/\chi_0^2)$ on the right-hand-side of~\eqref{Weq}. Now we can expand~\eqref{Weq} and perform the $q$ integral in essentially the same way as for the constant-field case. To zeroth order we find
\be\label{Wlead}
{\bf w}_{m,0}=\frac{\bm1}{(1+\rho J(\sigma))^m} \;,
\ee
where
\be
J(\sigma)=\int_{\sigma}^\infty\ud \sigma'\, F^2(\sigma') \;.
\ee
Note that it is the lower rather than upper integration limit that is given by $\sigma$, because our equations for the moments are integrated backwards in time. \eqref{Wlead} is, though, exactly what one would expect\footnote{Agreement between LL and the classical limit of kinetic equations in the LCF regime has been demonstrated in~\cite{Neitz:2013qba,NielPRE2018,Elkina:2010up}. In~\cite{Torgrimsson:2021wcj} it was also shown that the classical limit of the incoherent-product approximation, which we are using here, agrees with LL beyond the LCF regime.} from the exact solution to the classical Landau-Lifshitz (LL) equation~\cite{exactSolLL,Heintzmann:1972mn}. 

After expanding~\eqref{Weq} to the next-to-leading order, we find a partial differential equation which involves both $\partial_\sigma{\bf w}$ and $\partial_\rho{\bf w}$. We can make this an ordinary differential equation using the method of characteristics, i.e. we change variables from $\sigma$ and $\rho$ to $t=\sigma$ and 
\be\label{characteristicy}
y=\frac{1}{\rho}+J(\sigma) \;.
\ee 
After this change of variables we have an equation on the form
\be\label{firstOrderEq}
(y-J)^{m+2}\frac{\partial}{\partial t}\frac{{\bf w}_{m,1}}{(y-J)^{m+1}}=F^3{\bf R}(J,y) \;,
\ee
where the right-hand-side is a rational function of $J$ and $y$. It is now straightforward to integrate~\eqref{firstOrderEq}. We find for the zeroth, first and second moments,
\be\label{w01}
{\bf w}_{0,1}=
\int_{-\infty}^\infty\ud\sigma\frac{\rho F^3(\sigma)}{[1+\rho I(\sigma)]^2}
\frac{3}{2}
\begin{pmatrix}
0&\epsilon(\sigma)\\0&-\frac{5\sqrt{3}}{8}
\end{pmatrix}
\;,
\ee
\begin{widetext} 
\be\label{w11}
{\bf w}_{1,1}=\int_{-\infty}^\infty\ud\sigma\frac{\rho F^3(\sigma)}{[1+\rho I(\infty)]^2}
\begin{pmatrix}
\frac{55}{16\sqrt{3}}\left(\frac{1}{1+\rho I(\infty)}+\frac{2}{1+\rho I(\sigma)}\right) &
\frac{3\rho[I(\infty)-I(\sigma)]}{2[1+\rho I(\sigma)]^2}\epsilon(\sigma) \\
-\frac{3}{2[1+\rho I(\sigma)]}\epsilon(\sigma) & \frac{55}{16\sqrt{3}}\left(\frac{1}{1+\rho I(\infty)}+\frac{2}{1+\rho I(\sigma)}\right)-\frac{15\sqrt{3}[1+\rho I(\infty)]}{16[1+\rho I(\sigma)]^2} 
\end{pmatrix} \;,
\ee
\be\label{w21}
{\bf w}_{2,1}=\int_{-\infty}^\infty\ud\sigma\frac{\rho F^3(\sigma)}{[1+\rho I(\infty)]^3}
\begin{pmatrix}
\frac{55}{16\sqrt{3}}\left(\frac{3}{1+\rho I(\infty)}+\frac{4}{1+\rho I(\sigma)}\right) &
-\frac{3(1+2\rho I(\sigma)-\rho I(\infty))}{2[1+\rho I(\sigma)]^2}\epsilon(\sigma) \\
-\frac{3}{1+\rho I(\sigma)}\epsilon(\sigma) &
\frac{55}{16\sqrt{3}}\left(\frac{3}{1+\rho I(\infty)}+\frac{4}{1+\rho I(\sigma)}\right)-
\frac{15\sqrt{3}[1+\rho I(\infty)]}{16[1+\rho I(\sigma)]^2}
\end{pmatrix} \;,
\ee
\end{widetext}
where the components are such that ${\bf e}_0=\{1,0\}$ and $\hat{\bf B}_0=\{0,1\}$ in the ${\bf e}_0$-$\hat{\bf B}_0$ subspace we are considering in this section, and we have used
\be
J(-\infty)-J(\sigma)=\int_{-\infty}^\sigma\ud\sigma'\,F^2(\sigma'):=I(\sigma) \;.
\ee
For a constant field these results reduce to Eqs.~(92), (74) and (75) in~\cite{Torgrimsson:2023rqo}.

If the field oscillates in a symmetric fashion, then the off-diagonal elements of the above matrices tend to vanish upon integration over $\sigma$, due to the $\epsilon(\sigma)$ factors. This means, in particular, that there would be no Sokolov-Ternov effect\footnote{The standard Sokolov-Ternov result~\cite{Sokolov:1963zn,BaierSokolovTernov} for the induced polarization is actually for a case without RR, while here we consider induced polarization with RR.}, i.e. if the initial electron is unpolarized (${\bf N}=\{1,0\}$), then it will stay unpolarized. It also means that if we sum over the final spin states, then there is no dependence on the initial spin.
However, this does not mean that all spin effects are gone. If, for example, the initial state is completely polarized along the magnetic field axis, ${\bf N}=\{1,s\}$, where $s=\pm1$, then after interaction there is a nonzero probability to find the electron in the spin-flipped state ${\bf N}=\{1,-s\}$. So, if one prepares the initial electron to be spin up and the detector only measures electrons with spin up, then the resulting RR is different from the unpolarized case.

If we sum over the final spin, then, as in the constant-field case, we find that the standard deviation does not depend on the initial spin. We find
\be\label{standardLCF}
S^2=\frac{55\rho\chi_0}{16\sqrt{3}[1+\rho I(\infty)]^4}\int_{-\infty}^\infty\ud\sigma\, F^3(\sigma) \;.
\ee
\eqref{standardLCF} agrees with Eq.~(14) in~\cite{Blackburn:2023dey}. We make further comparison in Appendix~\ref{Comparison with literature}.

\begin{figure}
\includegraphics[width=\linewidth]{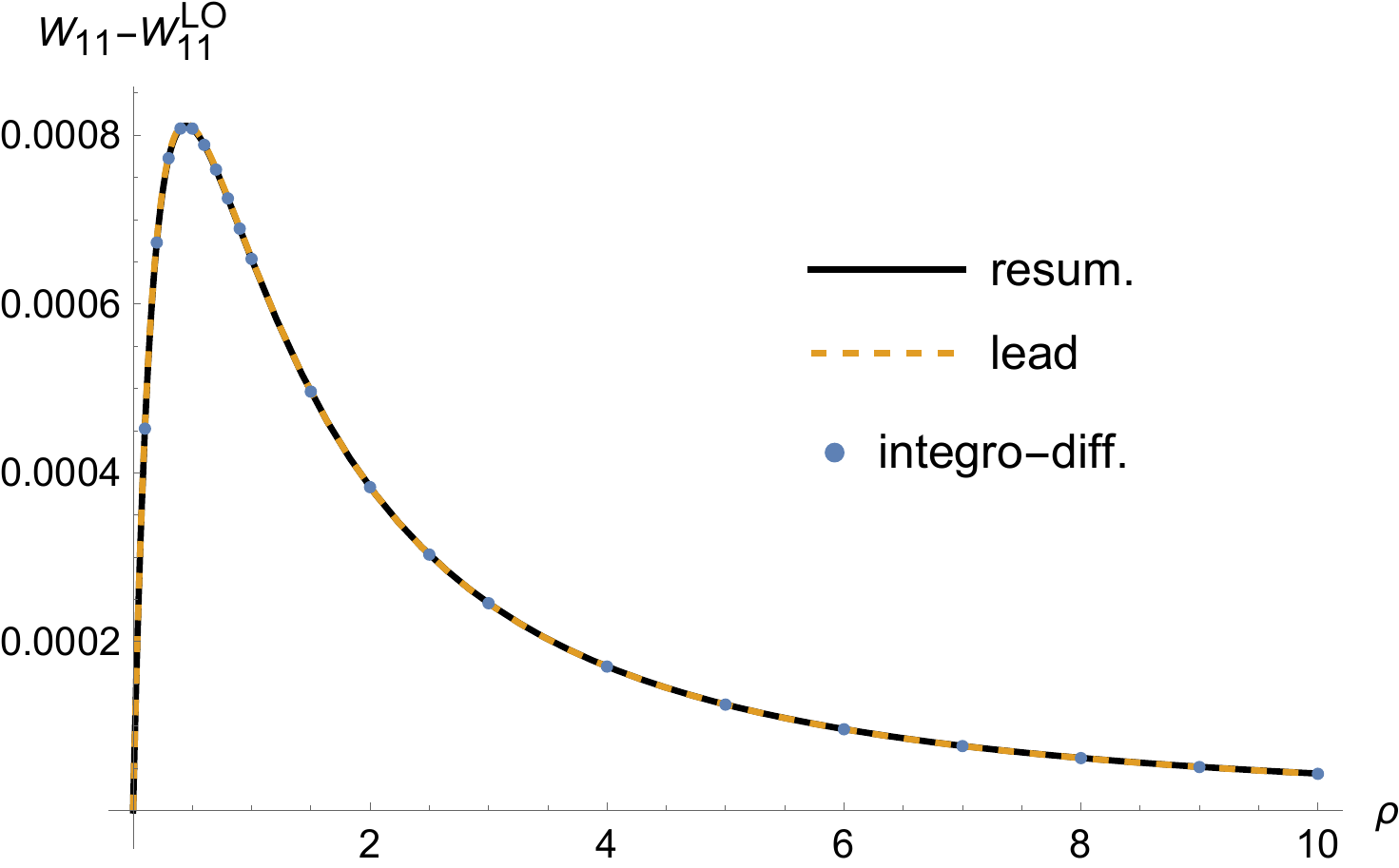}
\includegraphics[width=\linewidth]{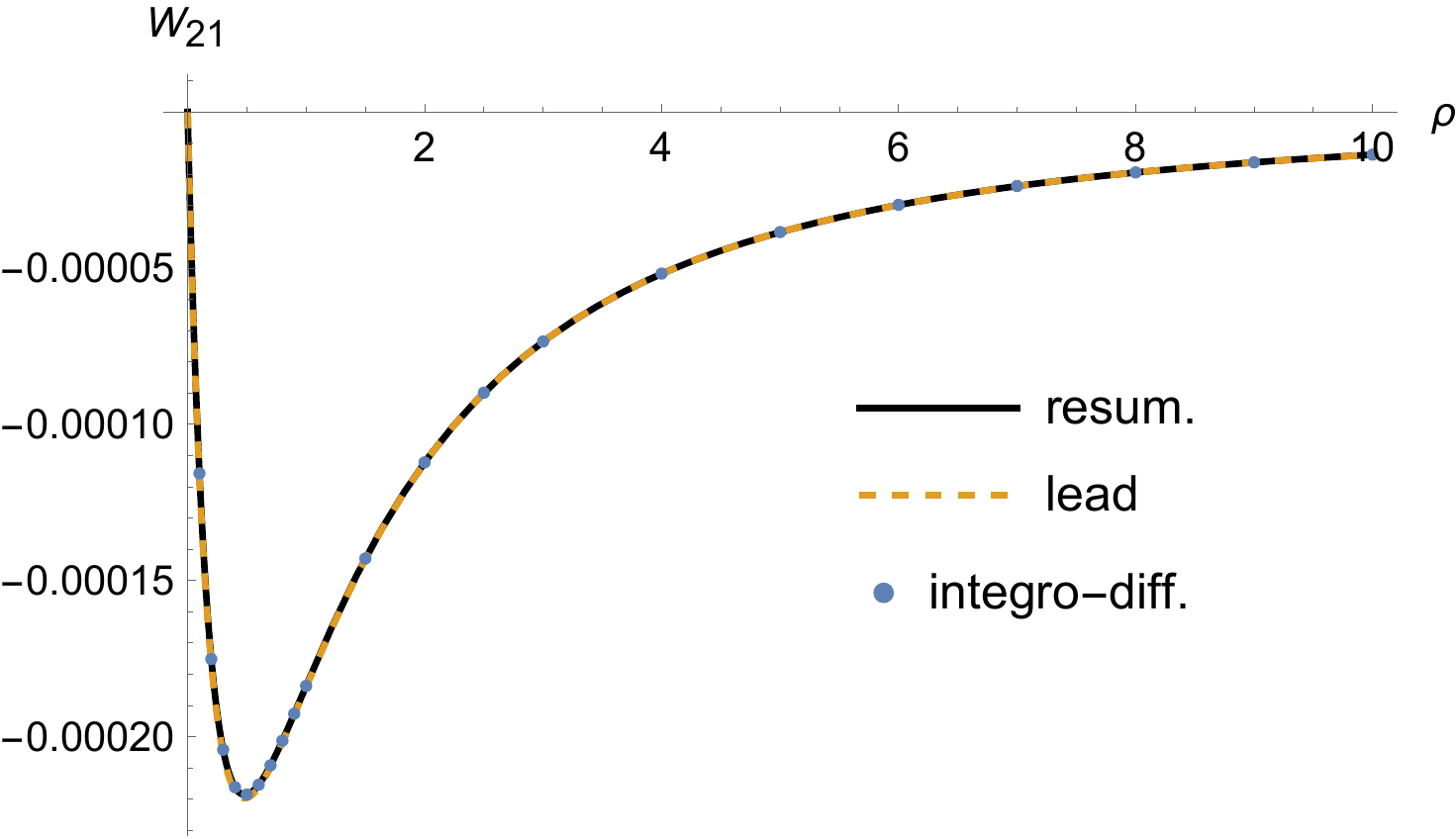}
\includegraphics[width=\linewidth]{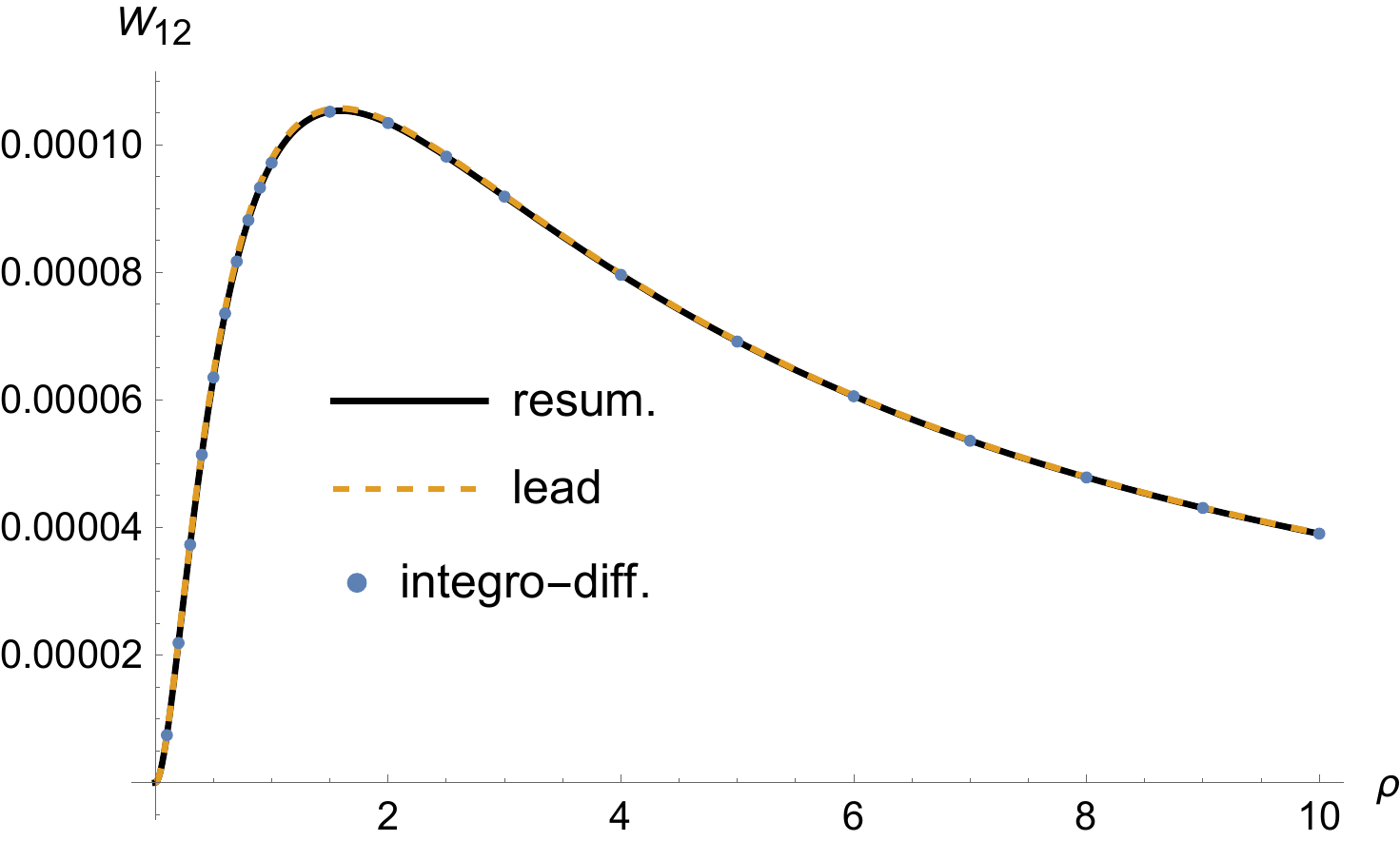}
\includegraphics[width=\linewidth]{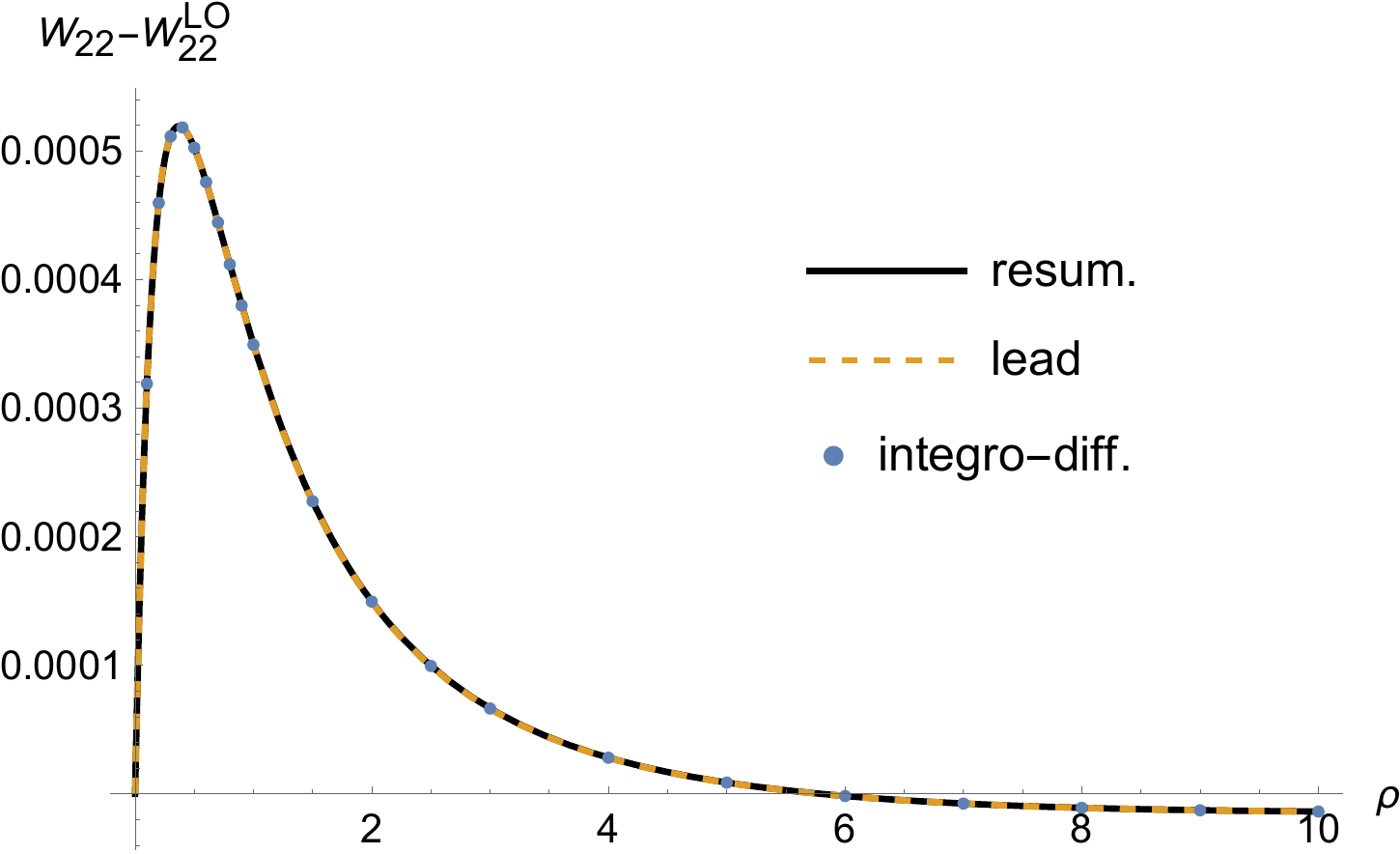}
\caption{The first moment, ${\bf W}(m=1)$ for a Sauter pulse and $\chi_0=10^{-3}$, for the components $W_{11}={\bf e}_0\cdot{\bf W}\cdot{\bf e}_0$, $W_{12}={\bf e}_0\cdot{\bf W}\cdot\hat{\bf B}_0$ etc. The ``lead'' curves show the analytical result for the leading order in~\eqref{w11}. The ``integro-diff.'' curves are numerical solutions to~\eqref{MnEq}. The ``resum.'' curves are obtained with the double resummation approach.}
\label{WsauterFig}
\end{figure}

\begin{figure}
\includegraphics[width=\linewidth]{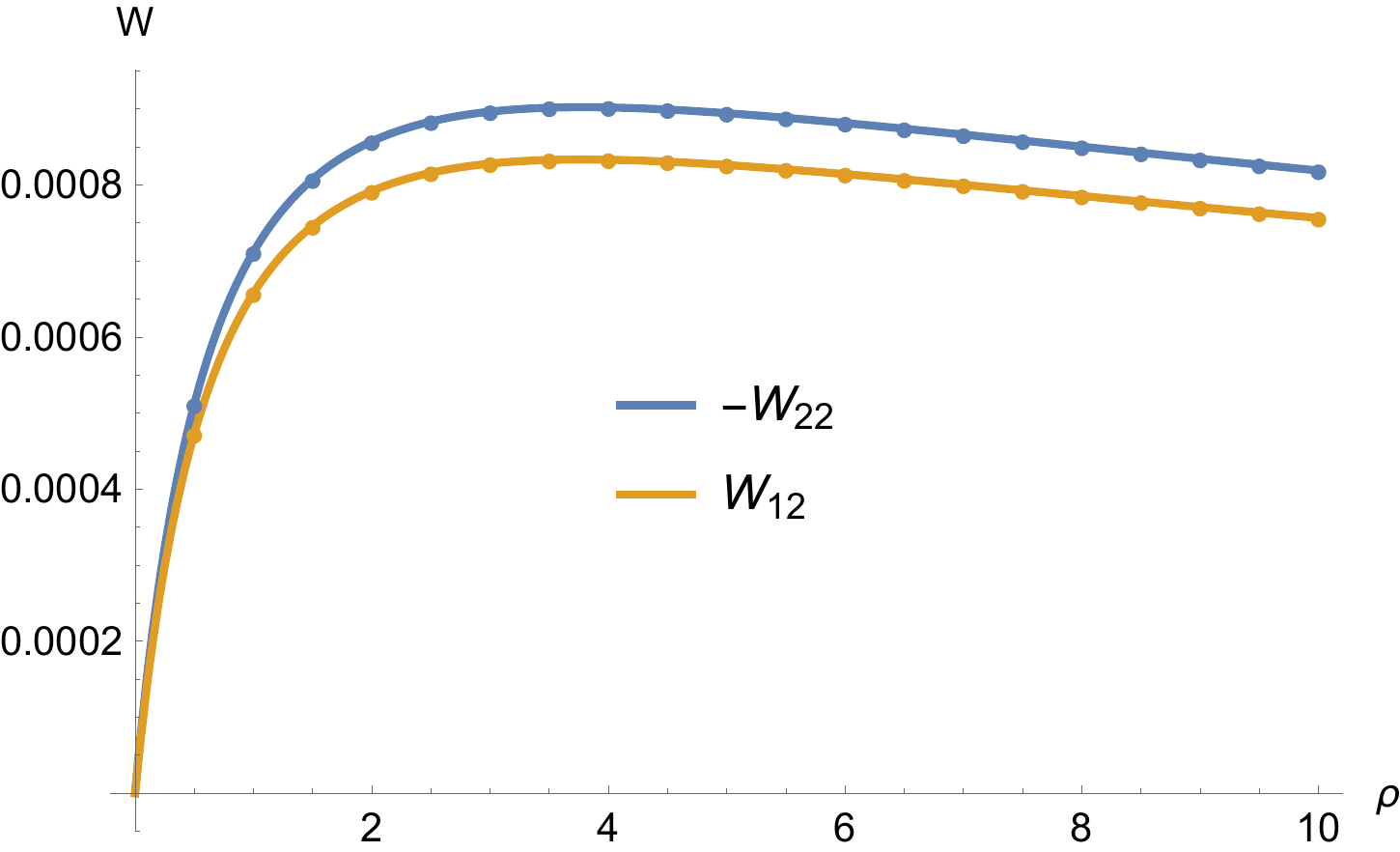}
\caption{The zeroth moment, ${\bf W}(m=0)$, for a Sauter pulse and $\chi_0=10^{-3}$. The solid lines are the leading order from~\eqref{w01}, and the dots are numerical solutions to~\eqref{MnEq}.}
\label{m0WsauterFig}
\end{figure}


We compare the analytical results in~\eqref{w01} and~\eqref{w11} with the exact, numerical result in Fig.~\ref{WsauterFig} and Fig.~\ref{m0WsauterFig} for $\chi_0=10^{-3}$. We have chosen such a small $\chi_0$ to make sure that the agreement is precise. We have obtained the numerical results in two very different ways.

\subsection{Direct numerical computation}

One way is to solve~\eqref{MnEq} by performing the integration over $\sigma$ using e.g. the midpoint method. At each point in $\sigma$ we make an interpolation function for the dependence on $\chi_0$ in the interval $0<\chi_0<\chi_{\rm max}$. 
Since we are often interested in small $\chi_{\rm max}$ and several different moments, rescaling and subtracting as follows helps with the precision and accuracy of the numerical computation.
First of all, we use~\eqref{Weq} instead of~\eqref{MnEq}, since ${\bf W}$ is on the same order of magnitude for all moments, while ${\bf M}$ decreases by orders of magnitude due to the overall $b_0^m$ scaling, so in terms of ${\bf W}$ we can work with the same accuracy for different $m$. We change integration variable from $q=\chi Y^3/(1+\chi Y^3)$ to $Y$. This makes the interval which gives significant contribution to the integral $\mathcal{O}(\chi^0)$. In contrast, in terms of $q$ the numerically important region would become smaller and smaller when we decrease $\chi$. Moreover, since we already know that the zeroth order is given by~\eqref{Wlead}, we subtract it in order to not waste precision and time computing it numerically. We therefore write
\be
{\bf W}=:{\bf w}_{m,0}+\chi_0\Delta{\bf W}
\ee
and rewrite~\eqref{Weq} as (omitting the $m$ subscript on ${\bf w}_{m,0}$)
\be\label{DeltaWeq}
\begin{split}
&\partial_\sigma\Delta{\bf W}=-\int_0^\infty\ud Y\mathcal{J}\bigg\{{\bf M}_L\cdot\Delta{\bf W}(b_0)\\
&+(1-q)^{m+1}{\bf M}_C\cdot\Delta{\bf W}([1-q]b_0)\\
&+\frac{1}{\chi_0}\bigg({\bf M}_L\cdot{\bf w}_0(b_0)+(1-q)^m{\bf M}_C\cdot{\bf w}_0([1-q]b_0)\\
&+\frac{1}{\mathcal{J}}[\mathcal{J} q{\bf M}_C]_\text{lead}\cdot\left[m{\bf w}_0+b_0\frac{\partial{\bf w}_0}{\partial b_0}\right]\bigg)
\bigg\} \;,
\end{split}
\ee
where $\mathcal{J}=3\chi Y^2(1+\chi Y^3)^2$ is the Jacobian for the change of integration variable, and
\be\label{MClead}
[\mathcal{J} q{\bf M}_C]_\text{lead}=-\frac{9\chi^2\rho}{2\chi_0^2}[Y^5\text{Ai}_1(Y^2)+2Y^3\text{Ai}'(Y^2)]{\bf 1}
\ee
is obtained by expanding $\mathcal{J} q{\bf M}_C$ to leading order in $\chi_0\ll1$ with $Y=\mathcal{O}(\chi_0^0)$. The term in the last row of~\eqref{DeltaWeq} is exactly equal to $-\partial_\sigma{\bf w}_0$. The reason for writing it as this $Y$ integral of~\eqref{MClead} is that it cancels the leading order of the third row, so the $\mathcal{O}(1/\chi_0)$ terms cancel already on the integrand level. \eqref{DeltaWeq} might look more complicated than~\eqref{Weq}, but~\eqref{DeltaWeq} is more convenient for numerical computation because the precision and accuracy goals of the integral are the same as that of $\Delta{\bf W}$. In contrast, if we instead solve~\eqref{Weq} in terms of ${\bf W}$, then for $\chi\ll1$ we lose precision when we subtract the leading order, ${\bf W}-{\bf w}_0$.

For the particular example of a Sauter pulse, $a_\mu=\{0,a,0,0\}$ with 
\be
a(\sigma)=a_0\tanh(\sigma)
\qquad
F(\sigma)=\text{sech}^2(\sigma) \;,
\ee
we changed variable from $\sigma$ to $t=\tanh(\sigma)$ and then integrated from $t=1$ to $t=-1$ with step size $\Delta t=1/100$. We solved~\eqref{DeltaWeq} for a couple of different values of $T=\alpha a_0=(3/2)(\rho_0/\chi_{\rm max})$ with $\rho_0=\{0.1,0.2,\dots 1,1.5,2,2.5,3,4\dots 10\}$ shown by dots in Fig.~\ref{WsauterFig}.

\subsection{Double resummation approach}\label{Double resummation approach}

The second method is to make a double resummation, where we express $\tilde{\bf M}$ in terms of its Taylor series in $\alpha$,
\be
\tilde{M}=\sum_{n=0}^\infty\tilde{\bf M}^{(n)} \;,
\ee 
where $\tilde{\bf M}^{(n)}$ is proportional to $\alpha^n$.
$\tilde{\bf M}^{(n)}$ is obtained from $\tilde{\bf M}^{(n-1)}$ using the recursive equation
\be\label{recursive}
\begin{split}
\tilde{\bf M}^{(n)}(b_0,\sigma)=&
\int_\sigma^\infty\ud\sigma'\int_0^1\!\ud q\bigg\{{\bf M}^L(\sigma')\!\cdot\!\tilde{\bf M}^{(n-1)}(b_0,\sigma')\\
&+{\bf M}^C(\sigma')\!\cdot\!\tilde{\bf M}^{(n-1)}\left([1-q]b_0,\sigma'\right)\bigg\} \;,
\end{split}
\ee
starting with
\be
\tilde{\bf M}^{(0)}(m,b_0)=b_0^m{\bf 1} \;.
\ee
The idea is to solve~\eqref{recursive} to obtain the first, say, $10$ or $15$ orders, and then resum the $\alpha$ expansion using Pad\'e approximants.
While~\eqref{recursive} can be solved by making a numerical interpolation function for the dependence on $\chi_0$ of each $\tilde{\bf M}^{(n)}$, we have instead solved it by making a second expansion, 
\be
\tilde{\bf M}^{(n)}(m)=\sum_k \chi_0^k\tilde{\bf M}^{(n,k)} \;.
\ee
For a Sauter pulse we can analytically obtain each coefficient $\tilde{\bf M}^{(n,k)}$ in this double expansion. We again change variable from $\sigma$ to $t=\tanh(\sigma)$, and then, after expanding in $\chi_0$, we find integrals on the form
\be\label{yintSauter}
\int_t^1\ud t'\, t^{\prime j}=\frac{1}{1+j}\left(1-t^{1+j}\right) \;,
\ee
with integer $j$. This simple formula is the main reason why we chose a Sauter field.
We have obtained the first $40$ orders in $\chi_0$ for each $\tilde{\bf M}^{(n)}$ up to $n=15$. We then resum the $\chi_0$ expansions for each $\tilde{\bf M}^{(n)}$ separately using the Borel-Pad\'e method, and finally we resum the $\alpha$ expansion using Pad\'e approximants. 
We need the full ``time'' $t$ dependence of ${\bf M}^{(n)}$ in order to calculate ${\bf M}^{(n+1)}$, so, in the intermediate steps in the calculation, $\tilde{\bf M}^{(n,k)}$ are functions of $t$, while they would be just constants for a constant field. However, after we have obtained all 10 or 15 orders we evaluate all these functions at $t=-1$. Thus, the actual resummations, using Borel-Pad\'e and Pad\'e, are essentially the same as in the constant field case. We therefore refer to~\cite{Torgrimsson:2021wcj,Torgrimsson:2021zob} for more details. The results are shown in black lines in Fig.~\ref{WsauterFig}.  
The precise agreement with the leading analytical $\mathcal{O}(\chi)$ results shows that in this particular regime it is unnecessary to resum the $\chi$ expansions.

\subsection{Higher moments and higher orders}\label{Higher moments}

\begin{figure}
\includegraphics[width=\linewidth]{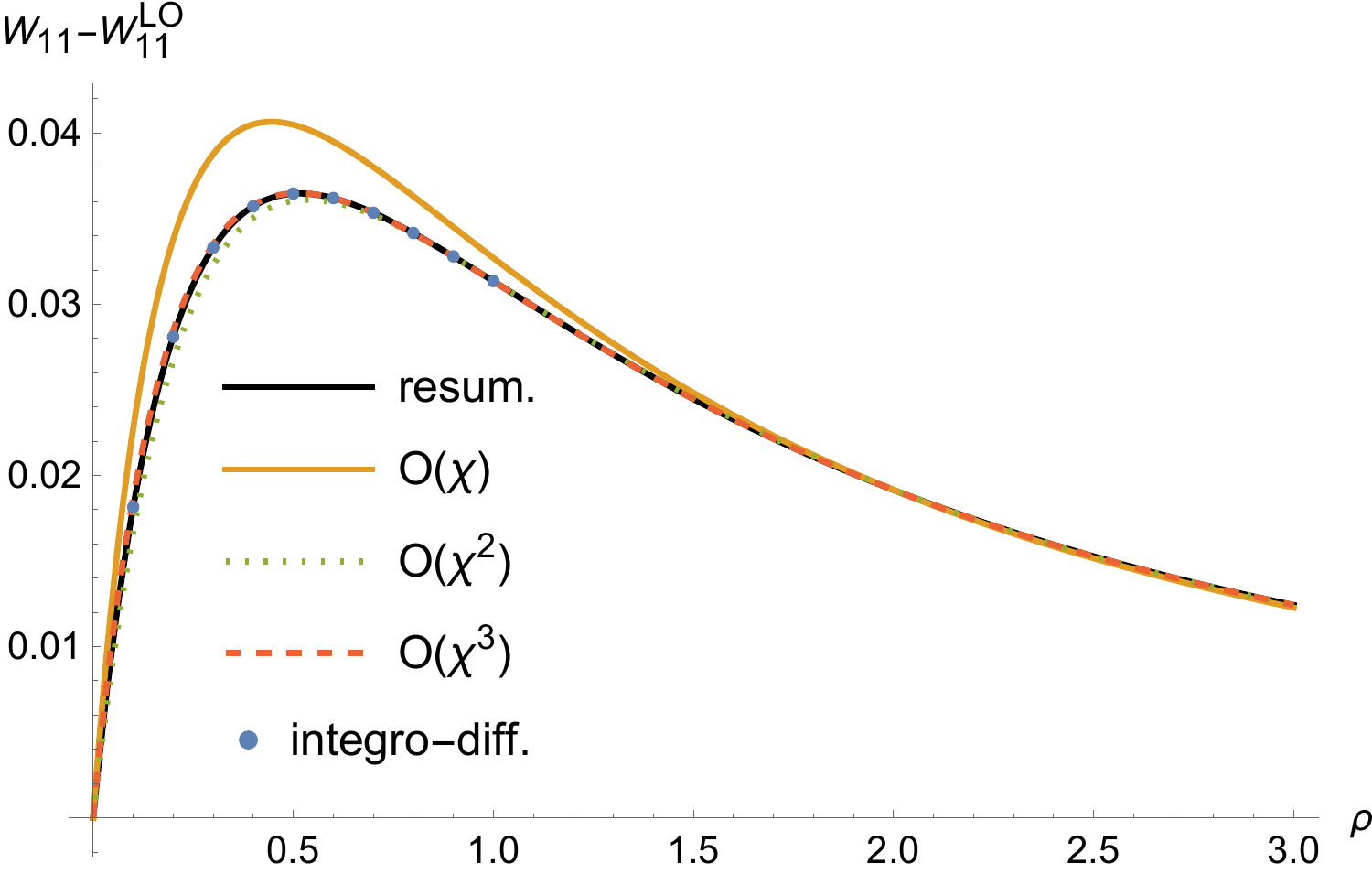}
\caption{Same as in Fig.~\ref{WsauterFig} but for $\chi_0=1/20$. $\mathcal{O}(\chi)$ is the leading order~\eqref{w11} (or~\eqref{omegam1}), $\mathcal{O}(\chi^2)$ is obtained by also including the NLO~\eqref{omegam2}, and $\mathcal{O}(\chi^3)$ includes the NNLO.}
\label{W3ordersFig}
\end{figure}

In this section we consider higher moments $m$ and the higher orders $\mathcal{O}(\chi^2)$ and $\mathcal{O}(\chi^3)$.
We first rescale ${\bf w}$ in~\eqref{Wsumw} as
\be\label{wtoomega}
{\bf w}_{m,k}=\frac{1}{\rho^{m+k}}{\bm\omega}_{m,k} \;.
\ee 
Then the zeroth order (the classical LL result) is
\be\label{omegam0y}
{\bm\omega}_{m,0}=\frac{\bf1}{y^m} \;,
\ee
where $y$ is given by~\eqref{characteristicy}. The generalization of~\eqref{w01}, \eqref{w11} and~\eqref{w21} to arbitrary moments is given by
\be\label{omegam1}
\begin{split}
{\bm\omega}_{m,1}&=
\int_\sigma^\infty\ud\varphi\, F^3(\varphi)\bigg\{\frac{55}{32\sqrt{3}}\frac{\ud^2{\bm\omega}_{m,0}}{\ud y^2}\\
&+\frac{1}{y-J(\varphi)}\begin{pmatrix}-\frac{55}{8\sqrt{3}}&\frac{3}{2}\\ \frac{3}{2}&-\frac{55}{8\sqrt{3}}\end{pmatrix}\cdot\frac{\ud{\bm\omega}_{m,0}}{\ud y}\\
&+\frac{1}{[y-J(\varphi)]^2}\frac{3}{2}\begin{pmatrix}0&1\\0&-\frac{5\sqrt{3}}{8}\end{pmatrix}\cdot{\bm\omega}_{m,0} \bigg\}
 \;.
\end{split}
\ee 
At $\mathcal{O}(\chi^2)$ we find
\begin{widetext}
\be\label{omegam2}
\begin{split}
{\bm\omega}_{m,2}&=
\int_\sigma^\infty\ud\varphi\, F^3\bigg\{\frac{55}{32\sqrt{3}}\frac{\ud^2{\bm\omega}_{m,1}}{\ud y^2}
+\frac{1}{y-J}\begin{pmatrix}-\frac{55}{8\sqrt{3}}&\frac{3}{2}\\ \frac{3}{2}&-\frac{55}{8\sqrt{3}}\end{pmatrix}\cdot\frac{\ud{\bm\omega}_{m,1}}{\ud y}
+\frac{1}{(y-J)^2}\frac{3}{2}\begin{pmatrix}0&1\\0&-\frac{5\sqrt{3}}{8}\end{pmatrix}\cdot{\bm\omega}_{m,1}\bigg\}\\
&+\int_\sigma^\infty\ud\varphi\, F^4\bigg\{\frac{7}{6}\frac{\ud^3{\bm\omega}_{m,0}}{\ud y^3}
+\frac{1}{y-J}\begin{pmatrix}-7&\frac{105\sqrt{3}}{64}\\ \frac{105\sqrt{3}}{64}&-7\end{pmatrix}\cdot\frac{\ud^2{\bm\omega}_{m,0}}{\ud y^2} \\
&\hspace{2.2cm}+\frac{1}{(y-J)^2}\begin{pmatrix}27&-\frac{105\sqrt{3}}{16}\\-\frac{315\sqrt{3}}{32}&21\end{pmatrix}\cdot\frac{\ud{\bm\omega}_{m,0}}{\ud y}
+\frac{1}{(y-J)^3}\begin{pmatrix}0&-\frac{315\sqrt{3}}{32}\\0&18\end{pmatrix}\cdot{\bm\omega}_{m,0}
\bigg\}
 \;.
\end{split}
\ee
\end{widetext}
We have also calculated a similar result for $\mathcal{O}(\chi^3)$.
We compare the above result with a numerical computation of~\eqref{DeltaWeq} in Fig.~\ref{W3ordersFig} and find good agreement. There is no problem to evaluate~\eqref{omegam2}, but, at least for a Sauter pulse, we found that it is much faster to first make a double expansion in $\chi\ll1$ and $\alpha$, as in the double-resummation approach in Sec.~\ref{Double resummation approach}, but keep the $\chi\ll1$ expansion un-resummed and only resum the $\alpha$ expansion (with $\rho=\mathcal{O}(\chi^0)$).  

\section{Mellin transform}\label{Mellin transform}

In this section we treat $m$ as a Mellin integration variable.
For $m\gg1$, only the first term in~\eqref{omegam1} and~\eqref{omegam2} contribute to leading order,
\be\label{omegakfromkm1}
{\bm\omega}_{m,k}\approx \frac{m^2}{y^2}\frac{55}{32\sqrt{3}}\int_\sigma^\infty\ud\varphi\, F^3{\bm\omega}_{m,k-1} \;,
\ee
for both $k=1$ and $k=2$. It is natural to guess that this formula will continue to hold for arbitrary $k$. Then we have
\be
\begin{split}
{\bm\omega}_{m,k}\approx&\frac{\bf1}{y^m}\left(\frac{m^2}{y^2}\frac{55}{32\sqrt{3}}\right)^k\int_\sigma^\infty\!\ud\varphi_k\, F^3(\varphi_k)\\
&\int_{\varphi_k}^\infty\!\ud\varphi_{k-1}\, F^3(\varphi_{k-1})\dots
\int_{\varphi_2}^\infty\!\ud\varphi_1\, F^3(\varphi_1)\\
=&\frac{\bf1}{y^m}\frac{1}{k!}\left(\frac{m^2}{y^2}\frac{55}{32\sqrt{3}}\int_\sigma^\infty\!\ud\varphi\, F^3\right)^k \;.
\end{split}
\ee
Summing over all orders in $\chi_0$ gives
\be\label{Wm2}
{\bf W}\approx\frac{\bf1}{(1+\rho J)^m}\exp\left\{\frac{\chi_0\lambda}{4}(1+\rho J)^2m^2\right\} \;,
\ee
where
\be\label{lambdaLCF}
\lambda=\frac{55\rho}{8\sqrt{3}[1+\rho J(\sigma)]^4}\int_\sigma^\infty\ud\sigma'\,F^3(\sigma') \;.
\ee
The reason for defining $\lambda$ like this will be explained in a moment.

Although we obtained~\eqref{Wm2} by summing over all orders in a $\chi_0\ll1$ expansion, it is better to think of the result as the leading order in a $\chi_0\ll1$ expansion where both $\rho=\mathcal{O}(1)$ and $\chi_0 m^2=\mathcal{O}(1)$. We will now rederive~\eqref{Wm2} (and hence confirm that~\eqref{omegakfromkm1} holds for all $k$) by assuming $\chi_0 m^2=\mathcal{O}(1)$ right from the start. We start with the ansatz
\be\label{Wtobarw}
{\bf W}=\frac{\bf1}{(1+\rho J)^m}\sum_{k=0}^\infty\bar{\bf w}_k(\sigma,\rho,\mu)\chi_0^{k/2} \;,
\ee 
where $\mu=m\sqrt{\chi_0}$. We plug this ansatz into~\eqref{Weq}, multiply both sides by $(1+\rho J)^m$, change variable from $q=\chi\gamma/(1+\chi\gamma)$ to $\gamma$ (same as before), and then expand the integrand in powers of $\sqrt{\chi_0}$ with $\rho=\mathcal{O}(1)$ and $\mu=\mathcal{O}(1)$. In the term proportional to ${\bf M}_C$, we have $\bar{\bf w}_k(\sigma,[1-q]\rho,\sqrt{1-q}\mu)$, which, after expanding in $\chi_0\ll1$, gives terms with derivatives of $\bar{\bf w}$ with respect to $\rho$ and $\mu$. The expansion of~\eqref{Weq} start at $\mathcal{O}(1/\sqrt{\chi_0})$, but both sides are automatically equal at this order. At $\mathcal{O}(\chi_0^0)$, we find
\be
\begin{split}
&\partial_\sigma\bar{\bf w}_0-\rho^2F^2\partial_\rho\bar{\bf w}_0-\frac{\mu\rho}{2}F^2\partial_\mu\bar{\bf w}_0\\
&=-\frac{55\mu^2\rho F^3}{32\sqrt{3}(1+\rho J)^2}\bar{\bf w}_0 \;.
\end{split}
\ee
We again solve this equation by changing variable from $\rho$ to $y$ as in~\eqref{characteristicy}. We find 
\be\label{barw0}
\bar{\bf w}_0={\bf 1}\exp\left\{\mu^2\frac{55}{32\sqrt{3}}\frac{y-J(\sigma)}{y^2}\int_\sigma^\infty F^3\right\} \;,
\ee   
which is the same as in~\eqref{Wm2}.

At $\mathcal{O}(\chi^{1/2})$ we find an equation which is solved by
\be
\bar{\bf w}_1=\bar{\bf w}_0\cdot\left(\frac{\mu}{\rho}{\bf H}_1^{(1)}+\frac{\mu^3}{\rho^2}{\bf H}_3^{(1)}\right) \;,
\ee
where
\be
\begin{split}
{\bf H}_1^{(1)}=\int_\sigma^\infty\ud\varphi\bigg\{&\frac{55F^3(\varphi)[5y-J(\varphi)]}{32\sqrt{3}y^2[y-J(\varphi)]}\begin{pmatrix}1&0\\0&1\end{pmatrix}\\
&-\frac{3F^3(\varphi)}{2y[y-J(\varphi)]}\begin{pmatrix}0&1\\1&0\end{pmatrix}\bigg\}
\end{split}
\ee
and
\be
{\bf H}_3^{(1)}=\int_\sigma^\infty\!\ud\varphi\bigg\{\frac{7F^4(\varphi)}{6y^3}-\frac{55^2F^3(\varphi)}{2^83y^4}\int_\varphi^\infty\! F^3\bigg\}\begin{pmatrix}1&0\\0&1\end{pmatrix} \;.
\ee

At $\mathcal{O}(\chi)$ we find
\be
\bar{\bf w}_2=\bar{\bf w}_0\cdot\left(\frac{1}{\rho}{\bf H}_0^{(2)}+\frac{\mu^2}{\rho^2}{\bf H}_2^{(2)}+\frac{\mu^4}{\rho^3}{\bf H}_4^{(2)}+\frac{\mu^6}{\rho^4}{\bf H}_6^{(2)}\right) \;,
\ee
where ${\bf H}_j^{(2)}$ depend on $\sigma$ and $y$ but not on $\mu$. With this ansatz for the dependence on $\mu$ the equations for ${\bf H}_j^{(2)}$ are easy to solve. For example,
\be
\begin{split}
&\partial_\sigma ({\bf H}_2^{(2)})_{1,1}=\frac{7F^4(3y-J)}{2y^3(y-J)}-\frac{3025F^3(7y-3J)}{1536y^4(y-J)}\int_\sigma^\infty F^3\\
&-\frac{55F^3(5y-J)}{32\sqrt{3}y^2(y-J)}({\bf H}_1^{(1)})_{1,1} +\frac{3F^3}{2y(y-J)}({\bf H}_1^{(1)})_{2,1}\\
&+\frac{55F^3}{16\sqrt{3}y}\partial_y({\bf H}_1^{(1)})_{1,1} \;,
\end{split}
\ee
where $(.)_{i,j}$ indicates components of the matrices, and the argument $\sigma$ for $F$, $J$ and ${\bf H}$ is left implicit. It is straightforward to solve these equations, but the resulting expressions for ${\bf H}_j^{(2)}$ are not particularly illuminating.  

Note that it was not necessary to assume that the moment order $m$ is an integer. In our approach, the equation for ${\bf W}(m)$ does not depend on other values of $m$. $m$ is just a parameter, and there is no problem to consider it to be an arbitrary continuous variable. In fact, in the following we will consider complex $m$ in order to perform an inverse Mellin transform. Note that~\eqref{Weq} allows us to find the moments without having to consider the spectrum ${\bf S}=\partial_x{\bf M}(b_0,x)$, where $x$ is related to the final momentum $p'_\mu$ as
\be\label{xdef}
kp'=(1-x)kp \;,
\ee
but if we have calculated ${\bf S}$ then we can obtain the same moments from
\be
{\bf W}(m)=\int_{-\infty}^1\ud x(1-x)^m{\bf S}=\int_0^\infty\ud v\,v^m{\bf S} \;,
\ee  
where we have changed variable from $x$ to $v=1-x$.
By letting $m$ be a continuous, complex parameter, this gives the Mellin transform of the spectrum. We can therefore obtain the spectrum from the moments by performing an inverse Mellin transform,
\be
{\bf S}=\frac{1}{v}\int_{-i\infty}^{i\infty}\frac{\ud m}{2\pi i}v^{-m}{\bf W}(m) \;.
\ee
We can obtain a $\chi_0\ll1$ expansion of ${\bf S}$ by plugging in the expansion in~\eqref{Wtobarw},
\be\label{Sfromwbar}
\begin{split}
{\bf S}=&\frac{1}{v\sqrt{\chi_0}}\int\frac{\ud\mu}{2\pi i}e^{a\mu^2-\mu L}\\
&\left(1+\sqrt{\chi_0}\left[\frac{\mu}{\rho}{\bf H}_1^{(1)}+\frac{\mu^3}{\rho^2}{\bf H}_3^{(1)}\right]+\chi_0[\dots]+\dots\right) \;,
\end{split}
\ee 
where $a>0$ is the coefficient of $\mu^2$ in the exponent of~\eqref{barw0} and
\be
L=\frac{1}{\sqrt{\chi_0}}\ln[(1+\rho J)v] \;.
\ee
We change variable from 
\be\label{mutonu}
\mu=\frac{L}{2a}+\frac{i}{\sqrt{a}}\nu
\ee 
to $\nu$ and integrate along the real axis, which corresponds to a contour parallel to the imaginary axis in the complex $m$ plane. As explained in~\cite{Torgrimsson:2023rqo}, it is natural to consider
\be
X=\frac{1}{\sqrt{\lambda\chi_0}}\left(\frac{1}{1+\rho J}-v\right)
\ee  
as an $\mathcal{O}(\chi^0)$ variable for the spectrum. This means, in particular,
\be
L=-\sqrt{\lambda}(1+\rho J)X+\mathcal{O}(\sqrt{\chi_0}) \;.
\ee
Thus, after performing the Gaussian integrals, we expand the result in powers of $\chi_0$ with $X$ rather $v$ (or x) as $\mathcal{O}(\chi^0)$. We find
\be
\begin{split}
{\bf S}=&\frac{e^{-X^2}}{\sqrt{\pi\lambda\chi_0}}\bigg\{{\bf 1}+\frac{\sqrt{\chi_0}X}{\sqrt{\lambda}}\bigg[\lambda(1-X^2)(1+\rho J){\bf 1}\\
&-\frac{2{\bf H}_1^{(1)}}{\rho(1+\rho J)}
+\frac{12-8X^2}{\lambda\rho^2(1+\rho J)^3}{\bf H}_3^{(1)}\bigg]+\mathcal{O}(\chi_0)\bigg\} \;.
\end{split}
\ee  
For a constant field we can compare with the results in~\cite{Torgrimsson:2023rqo}. We can immediately see that the leading order is the same. The $\mathcal{O}(\sqrt{\chi_0})$ term agrees with Eq.~(111) and (112) in~\cite{Torgrimsson:2023rqo}. We have also checked that the $\mathcal{O}(\chi_0)$ term agrees the corresponding constant field result, which was used for plots in~\cite{Torgrimsson:2023rqo} but not explicitly shown. We explain how to obtain the spectrum for a non-constant field with the same approach as in~\cite{Torgrimsson:2023rqo} in Appendix~\ref{LCF of spectrum}.

One could also perform the inverse Mellin transform numerically. If one still makes the change of variable in~\eqref{mutonu}, then the integrand behaves roughly as $e^{-\nu^2}$ close to $\nu=0$. Since $e^{-4^2}=\mathcal{O}(10^{-7})$, one could expect to find a good precision by integrating from $\nu=-4$ to $\nu=4$. One could then choose a set of points in this interval, solve~\eqref{Weq} at each of these points, and then make an interpolation function for ${\bf W}(\nu)$. However, we leave that for future studies.

\section{Maximum entropy}\label{Maximum entropy}

\begin{figure}
\includegraphics[width=\linewidth]{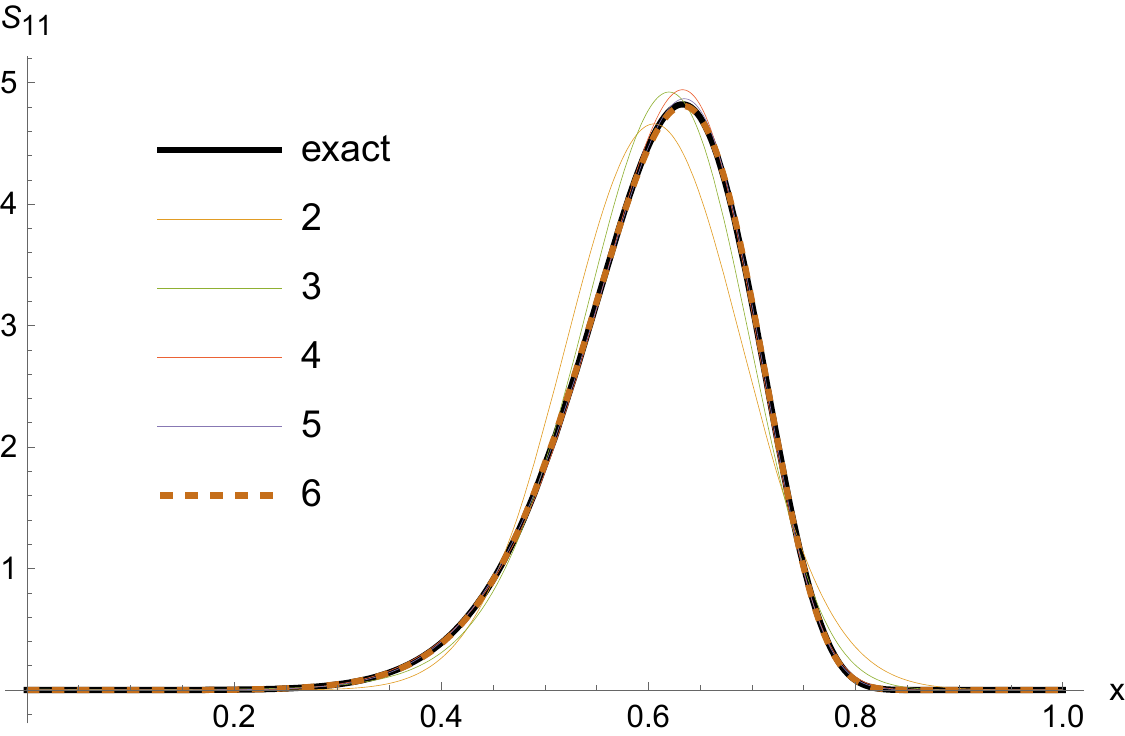}
\caption{The spectrum $S_{11}={\bf e}_0\cdot{\bf S}\cdot{\bf e}_0$ for a constant field with $\chi=0.1$ and $u:=(2/3)\chi\alpha a_0\Delta\phi=2$. The black line shows the numerical result obtained in~\cite{Torgrimsson:2023rqo} by solving directly the equation for the spectrum (or rather the cumulative function, which is then differentiated). The other lines show the maximum-entropy distribution $S_N$ in~\eqref{SNexp} with $N=2,3\dots 6$.}
\label{maxEntFig}
\end{figure}

\begin{figure}
\includegraphics[width=\linewidth]{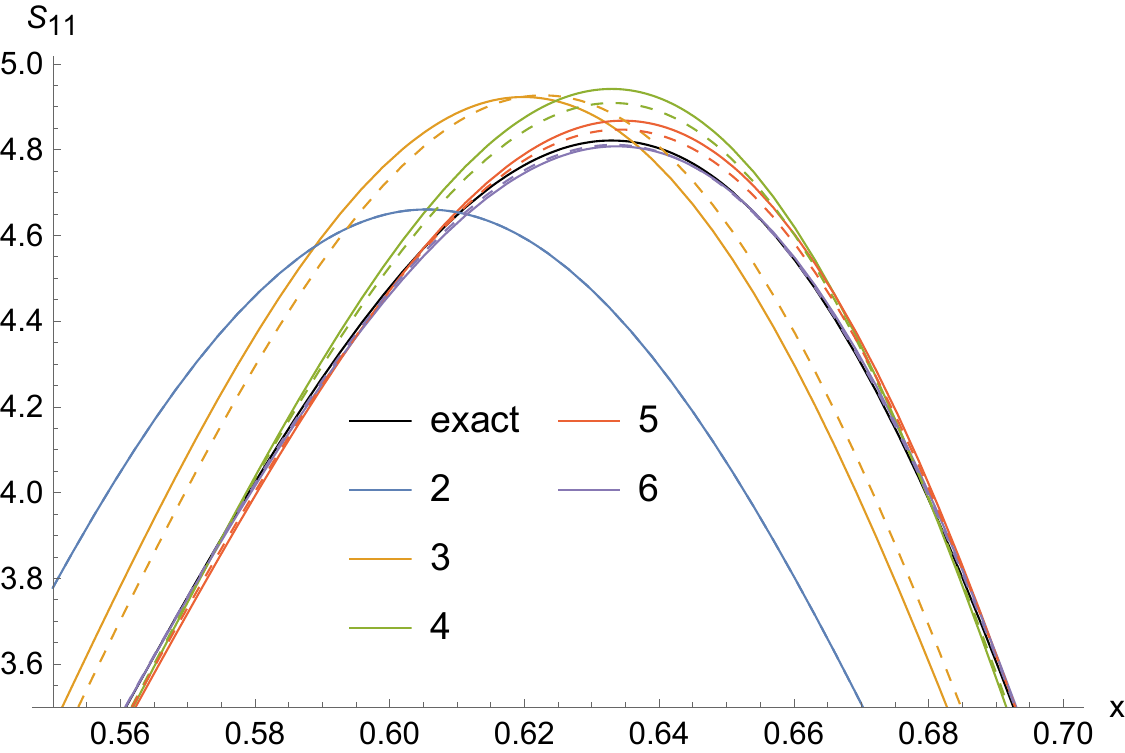}
\caption{The solid lines are the same as in Fig.~\ref{maxEntFig}, the dashed lines show~\eqref{MNerf}. The solid and dashed lines for $N=2$ coincide.}
\label{maxEntErfFig}
\end{figure}

In Sec.~\ref{Mellin transform} we showed how to obtain the spectrum by treating the moment order $m$ as a Mellin variable, and then integrating over $m$ in the complex plane. In this section we will show another way, where we obtain the spectrum from a handful of moments by using the principle of maximum entropy~\cite{Jaynes57,Jaynes68,maxent84}. The idea is to approximate the true spectrum $S_{11}={\bf e}_0\cdot{\bf S}\cdot{\bf e}_0$ (we average and sum over the initial and final spin for simplicity) with a distribution $S_N$ which maximizes the entropy subject to the constraints on the moments,
\be\label{SNmoments}
\int\ud x(1-x)^m S_N\overset{!}{=}{\bf e}_0\cdot{\bf W}(m)\cdot{\bf e}_0
\ee 
for $m=0,1,2\dots,N$. The function that satisfies these constraints and maximizes the entropy is given by~\cite{Jaynes57,Jaynes68,maxent84}
\be\label{SNexp}
S_N=p(x)\exp\left\{-\sum_{n=0}^N\lambda_n x^n\right\} \;,
\ee
where $\lambda_n$ are constants determined by~\eqref{SNmoments}. For the case we focus on here we expect $S(x)$ to be not too different from a Gaussian, we therefore choose the prior distribution~\cite{Jaynes68} $p(x)=1$ (choosing $p(x)=a\exp[-c(x-b)^2]$ with some constants $a$, $b$ and $c$ gives the same result as it just corresponds to a shift in $\lambda_0$, $\lambda_1$ and $\lambda_2$).
In Fig.~\ref{maxEntFig}, we compare the results with the exact numerical result obtained in~\cite{Torgrimsson:2023rqo} for a constant field. At $N=5$ or $N=6$ we have a good agreement. This is good news because each moment only takes a couple of minutes to compute, and solving~\eqref{SNmoments} is also very fast. So this maximum-entropy approach is much faster than a direct computation of the spectrum as in~\cite{Torgrimsson:2023rqo}.  

The principle of maximum entropy alone does not imply that having a polynomial in the exponent~\eqref{SNexp} is necessarily the best way to approximate the spectrum. There is a polynomial in the exponent simply because we have chosen to impose the constraints~\eqref{SNmoments} for integer moments. If one instead chooses a set of fractional~\cite{MaxEntFractional} or negative~\cite{MaxEntNegative} powers $m$ (or the expectation value of some more nontrivial function), then one has a sum of such terms in the exponent. So, one can try to figure out what sort of $x$ dependence to expect and then let that guide the choice of $m$'s in order to find convergence with even fewer $m$'s.
We leave such investigations for the future and instead consider the following approach. 

The maximum entropy is expected to give the best estimate given a set of moments but assuming no other information. However, in our case we do have some additional information provided by the $\chi\ll1$ expansion in~\eqref{Sansatz}. We consider again a constant field. The results for ${\bm\rho}_1$ up to ${\bm\rho}_5$ were calculated for~\cite{Torgrimsson:2023rqo}, but only ${\bm\rho}_1$ was explicitly presented. What is relevant here is that 
\be
\begin{split}
{\bm\rho}_1(u,X)&=\sum_{n=0}^1{\bm\rho}_{1,2n+1}(u)X^{2n+1}\\
{\bm\rho}_2(u,X)&=\sum_{n=0}^3{\bm\rho}_{2,2n}(u)X^{2n}\\
{\bm\rho}_3(u,X)&=\sum_{n=0}^4{\bm\rho}_{1,2n+1}(u)X^{2n+1}\\
{\bm\rho}_4(u,X)&=\sum_{n=0}^6{\bm\rho}_{1,2n}(u)X^{2n}\\
{\bm\rho}_5(u,X)&=\sum_{n=0}^7{\bm\rho}_{1,2n+1}(u)X^{2n+1} \;.
\end{split}
\ee   
If we just exponentiate these terms,
\be\label{rhoToExp}
\begin{split}
&{\bf e}_0\cdot\sum_{n=0}^{5}{\bm\rho}_n(u,X)\chi^{n/2}\cdot{\bf e}_0\\
&=:\exp\left\{\sum_{n=1}^5 E_n(u,X)\chi^{n/2}\right\}+\mathcal{O}(\chi^{6/2}) \;,
\end{split}
\ee
where the coefficients $E_n$ are defined by taking the logarithm of the left-hand side and expanding to $\mathcal{O}(\chi^{5/2})$, then we find that $E_1$ contains terms proportional to either $X$ or $X^3$; $E_2$ contains $X^0$, $X^2$ and $X^4$; $E_3$ contains odd powers from $X$ to $X^5$; $E_4$ contains even powers from $X^0$ to $X^6$, and $E_5$ contains odd powers from $X$ to $X^7$. Thus, if $N(.)$ counts the number of terms in the $X$ polynomial,
\be\label{Nrho}
N(\rho_1,\rho_2,\rho_3,\rho_4,\rho_5)=(2,4,5,7,8)
\ee
while
\be\label{NE}
N(E_1,E_2,E_3,E_4,E_5)=(2,3,3,4,4)
\ee
We see that, at the very least, the information stored in $(\rho_1,\rho_2,\rho_3,\rho_4,\rho_5)$ can be more compactly stored in $(E_1,E_2,E_3,E_4,E_5)$. We can make it even a bit more compact in terms of $R_n$ defined by
\be\label{erfRsum}
\begin{split}
&\frac{e^{-X^2}}{\sqrt{\pi}}{\bf e}_0\cdot\sum_{n=0}^{5}{\bm\rho}_n(u,X)\chi^{n/2}\cdot{\bf e}_0\\
&=:\frac{1}{2}\frac{\partial}{\partial X}\text{erf}\left\{X+\sum_{n=1}^5 R_n(u,X)\chi^{n/2}\right\}+\mathcal{O}(\chi^{6/2}) \;.
\end{split}
\ee
The inspiration for this ansatz has nothing to do with the maximum entropy principle. Instead it comes from noting that the $\chi\ll1$ expansion of the cumulative function can be expressed as
\be
\begin{split}
{\bf M}&=\frac{1}{2}(1+\text{erf}(X)){\bf 1}-\frac{e^{-X^2}}{\sqrt{\pi}}\sum_{n=1}^\infty{\bf h}_n(u,\chi) H_{n-1}(X) \\
&=\sum_{n=0}^\infty{\bf h}_n(u,\chi)\left(-\frac{\ud}{\ud X}\right)^n\frac{1}{2}(1+\text{erf}(X))\;,
\end{split}
\ee  
where $H$ are Hermite polynomials and ${\bf h}_0={\bf 1}$, suggesting that one interpret it as the Taylor expansion of $\text{erf}(X+\dots)$. The actual justification for~\eqref{erfRsum} is that it does indeed lead to a more compact result,
\be
N(R_1,R_2,R_3,R_4,R_5)=(2,2,3,3,4) \;.
\ee 

In~\eqref{rhoToExp} we have the exponential of a polynomial in $x$, but it is not a proof of~\eqref{SNexp}. In~\eqref{rhoToExp} there are no undetermined quantities, i.e. one can evaluate it without matching with~\eqref{SNmoments}. However, \eqref{rhoToExp} is not a replacement of~\eqref{SNexp} either, because~\eqref{rhoToExp} only provides a partial resummation of the $\chi\ll1$ series. To obtain a full resummation one would need to resum $\sum E_n\chi^{n/2}$. \eqref{erfRsum}, too, only give a partial resummation. Thus, if $\chi$ is small but not very small, one can expect the $\chi\ll1$ expansions to break down regardless of whether one uses the expansion in terms of ${\bm\rho}_n$, $E_n$ or $R_n$. We instead take~\eqref{rhoToExp} as motivation for choosing the prior distribution~\cite{Jaynes68} such that we have~\eqref{SNexp} with $p(x)=1$. 

Similarly, we have also taken~\eqref{erfRsum} as motivation for matching the moments onto
\be\label{MNerf}
M_N=\frac{1}{2}\left[1+\text{erf}\left(\sum_{n=0}^{N-1}\kappa_n x^n\right)\right] \;,
\ee 
where the coefficients $\kappa_n$ are obtained by
\be
\int_0^1\ud x\, m(1-x)^{m-1}M_N\overset{!}{=}{\bf e}_0\cdot{\bf W}(m)\cdot{\bf e}_0 \;.
\ee 
This provides an alternative to the maximum-entropy fit~\eqref{SNexp}.  \eqref{SNexp} and~\eqref{MNerf}, obtained using the same set of moments, are compared in Fig.~\ref{maxEntErfFig}. We see that~\eqref{MNerf} provides a slight improvement compared to~\eqref{SNexp}. More importantly, having two different ways of obtaining the spectrum from the moments allows us to estimate the error in other cases when we do not already have the exact spectrum.

In Fig.~\ref{maxEntFig} we have chosen $u$ sufficiently large so that the peak is well separated from $x=0$. As seen in Fig.~1 or 3 in~\cite{Torgrimsson:2023rqo}, for small $u$ we instead have a sharp peak at $x=0$, which does not look like a Gaussian even to a first approximation. After trying to anyway use~\eqref{SNexp} with $p(x)=1$ for one such example, it looks like one can see that there is a peak at $x=0$, but the convergence is much slower. So, even if we did not have access to the exact spectrum, we would be able to tell that~\eqref{SNexp} with $p(x)=1$ is not as good in this case.  
This suggests that one should try to find another prior distribution $p(x)$ which captures the sharp peak to some rough first approximation, and/or choose to match with different moments. We leave that for future studies.

\section{Beyond LCF}\label{Beyond LCF}

In this section we will generalize the results in Sec.~\ref{LCF of moments} beyond the LCF regime. The incoherent-product approximation is still valid even if $a_0$ is not large, provided the pulse is long enough.  

We begin by noting that the longitudinal-momentum fraction is small, $q=kl/kp=\mathcal{O}(b_0)$, where $l$ is the photon momentum. We can therefore Taylor expand the moments
\be\label{MbTaylor}
\tilde{\bf M}([1-q]b_0)\approx\tilde{\bf M}(b_0)-qb_0\frac{\partial\tilde{\bf M}}{\partial b_0}+(qb_0)^2\frac{\partial^2\tilde{\bf M}}{\partial b_0^2} \;.
\ee
Although $b_0\ll1$, we cannot expand $\tilde{\bf M}(b_0)$ since $b_0$ appears in $\rho=\mathcal{O}(1)$. We plug this expansion into~\eqref{MnEq}, change variable from
\be
q=\frac{b_0\gamma}{1+b_0\gamma}
\ee 
to $\gamma=r/b_0$, expand the integrand (but not $\tilde{\bf M}(b_0)$) in $b_0\ll1$ with $\gamma$ and $\theta$ independent of $b_0$, and perform the resulting $\gamma$ integrals. Since we now consider more general regimes, ${\bf M}_{L,C}$ can in general not be expressed in terms of either Airy functions (for LCF) or Bessel functions (for LMF), instead we have $\theta$ integrals as in Eq.~(19) in~\cite{Torgrimsson:2023rqo} (see~\cite{Dinu:2019pau} for the other components of ${\bf M}_C$ for photon polarization). The definitions of ${\bf R}$, ${\bf V}$ etc. below can be found in~\cite{Torgrimsson:2023rqo}.
In the derivation of the Mueller matrices two lightfront-time variables appear as $\sigma=(\phi_2+\phi_1)/2$ and $\theta=\phi_2-\phi_1$, where $\phi_{1,2}$ are lightfront time variables for the amplitude and its complex conjugate. In the LCF regime we have $\theta=\mathcal{O}(1/a_0)$, so then when expanding in $1/a_0\ll1$ the probabilities localizes, i.e. can be expressed in terms of the field and its first derivatives evaluated at $\theta=0$. Thus, in the results below where the $\theta$ integrals receive significant contributions for regions where $\theta$ is not small, we have clear nonlocal, beyond-LCF effects. 

Performing the $\gamma$ integral before the $\theta$ integral gives
\be\label{gammaIntn}
\int_0^\infty\ud\gamma\,\gamma^n\exp\left(\frac{i}{2}\Theta\gamma\right)=n!\left(-\frac{i}{2}\Theta\right)^{-1-n} \;,
\ee 
where $\Theta=\theta M^2$ and $M^2(\theta,\sigma)$ is the effective mass~\cite{Kibble:1975vz}. The resulting $\theta$ integrals are now either symmetric or antisymmetric in $\theta$. In the antisymmetric case the $\theta$ integral is equal to $-i\pi$ times the residue at $\theta=0$ (the $\theta$ contour goes above the pole at $\theta=0$), which means that the probability localizes even though there is no small or large parameter that forces it to be so.

We consider first the matrix multiplying $\tilde{\bf M}(b_0)$,
\be
\left[\int_0^1\ud q({\bf M}_L+{\bf M}_C)\right]\cdot\tilde{\bf M}(b_0):={\bf m}_0\cdot\tilde{\bf M} \;.
\ee
Unitarity ensures that some of the components in the above sum cancel,
\be\label{RLplusRC}
{\bf R}_L+{\bf R}_C=\begin{pmatrix}0 & {\bf R}_1^C-{\bf R}_0^C\\
{\bm 0} & {\bf R}_{01}^C-\langle\mathbb{R}^C\rangle{\bm 1}+{\bf R}_{01}^{\rm rot} \end{pmatrix} \;.
\ee
We have
\be
{\bf R}_1^C-{\bf R}_0^C=\frac{q^2}{s}{\bf V} \;,
\ee
and using~\eqref{gammaIntn} with $n=2$, we find
\be
{\bf e}_0\cdot{\bf m}_0=\frac{8\alpha b_0^2}{\pi}\int\ud\theta\frac{\bf V}{\theta\Theta^3} \;.
\ee
The integrand is odd since $\Theta(-\theta)=-\Theta(\theta)$ and ${\bf V}(-\theta)=-{\bf V}(\theta)$, so, if we integrate along the real axis from $-\infty$ to $-\epsilon<0$, then clockwise around a semi-circle to $+\epsilon$, and finally along the real axis from $\epsilon$ to $+\infty$, the two contributions from the real axis cancel, and in the limit $\epsilon\to0$ we are left with $-i\pi$ times the residue at $\theta=0$. We find
\be\label{e0m0}
{\bf e}_0\cdot{\bf m}_0=\alpha b_0^2 i{\bm\sigma}_2\cdot\left[{\bf a}^{\prime2}{\bf a}'-\frac{1}{6}{\bf a}'''\right] \;,
\ee 
where we have a frame where ${\bf a}=a_1{\bf e}_1+a_2{\bf e}_2$ and 
\be
{\bm\sigma}_2=i({\bf e}_2{\bf e}_1-{\bf e}_1{\bf e}_2)
\ee
is one of the Pauli matrices,  
and the field is evaluated at $\sigma$, so this is a local term. If the field is linearly polarized with the electric field along the $x$ axis, ${\bf a}\propto{\bf e}_1$, then ${\bf e}_0\cdot{\bf m}_0\propto{\bf e}_2$ is parallel or antiparallel to the magnetic field.  

Next we consider the diagonal terms in ${\bf m}_0$, which come from the diagonal terms of $ {\bf R}_{01}^C-\langle\mathbb{R}^C\rangle{\bm 1}$ in~\eqref{RLplusRC}. Using again~\eqref{gammaIntn} with $n=2$ gives
\be\label{m0diag}
{\bf m}_0|_{\rm diag.}=\frac{2}{\pi}\alpha b_0^2\int\frac{\ud\theta}{\theta\Theta^3}\left(\Delta a^2[{\bf e}_1{\bf e}_1+{\bf e}_2{\bf e}_2]-4{\bf e}_3{\bf e}_3\right) \;,
\ee
where
\be
\Delta a^2=\left[{\bf a}\left(\sigma+\frac{\theta}{2}\right)-{\bf a}\left(\sigma-\frac{\theta}{2}\right)\right]^2 \;.
\ee
For this term we have a symmetric $\theta$ integrand, so the integral is not simply given by the residue at $\theta=0$. This is therefore a nonlocal term.  

${\bf m}_0$ also contains off-diagonal terms due to ${\bf R}_{01}^C$ and ${\bf R}_{01}^{\rm rot}$ in~\eqref{RLplusRC}. For a linearly polarized field, ${\bf a}\propto{\bf e}_1$, these off-diagonal terms are proportional to ${\bf e}_1{\bf e}_3$ or ${\bf e}_3{\bf e}_1$ and make the spin components in the $\hat{\bf E}$-$\hat{\bf k}$ plane (${\bf e}_1$-${\bf e}_3$ plane) rotate. We do not need to consider these terms if we consider initial and final spins along the magnetic field direction. For a symmetrically oscillating field with circular polarization those off-diagonal terms average out and we are instead left with a term proportional to $\sigma_2$, which leads to spin rotation in the plane perpendicular to the propagation direction $\hat{\bf k}={\bf e}_3$. In this case we can omit this off-diagonal term by considering initial and final spin parallel to $\hat{\bf k}$. We will explain one way of dealing with fast spin rotation in Sec.~\ref{Spin perpendicular to magnetic field}. For now, we consider spin components that do not rotate.  
 
We turn to the matrices, ${\bf m}_1$ and ${\bf m}_2$, which multiply $\partial_{b_0}\tilde{\bf M}$ and $\partial_{b_0}^2\tilde{\bf M}$ in~\eqref{MbTaylor}. Only ${\bf M}_C$ contributes to these terms since they come from expanding the second term in~\eqref{MnEq}. For the diagonal elements we find for the leading and next-to-leading order
\be\label{m1diag}
{\bf m}_1|_{\rm diag.}=\frac{2}{3}\alpha b_0{\bf a}^{\prime2}{\bm1}_4+\frac{12}{\pi}\alpha b_0^2{\bm1}_4\int\frac{\ud\theta}{\theta\Theta^3}[2+\Delta a^2] 
\ee
and
\be\label{m2diag}
{\bf m}_2|_{\rm diag.}=-\frac{4}{\pi}\alpha b_0^2{\bm1}_4\int\frac{\ud\theta}{\theta\Theta^3}[2+\Delta a^2] \;,
\ee
where ${\bm1}_4=\sum_{j=0}^4{\bf e}_j{\bf e}_j$.

As for ${\bf m}_0$, we can again omit the off-diagonal terms proportional to ${\bf e}_i{\bf e}_j$ if we consider spin components that do not rotate.
The off-diagonal terms proportional to ${\bf e}_0{\bf e}_i$ and ${\bf e}_i{\bf e}_0$, though, cannot be omitted. We find
\be\label{e0m1}
{\bf e}_0\cdot{\bf m}_1={\bf m}_1\cdot{\bf e}_0={\bf e}_0\cdot{\bf m}_0+\frac{2}{3}\alpha b_0^2[a_1'a_2''-a_2'a_1'']\hat{\bf k} \;,
\ee
where ${\bf e}_0\cdot{\bf m}_0$ is given by~\eqref{e0m0}. ${\bf e}_0\cdot{\bf m}_2$ and ${\bf m}_2\cdot{\bf e}_0$ are higher order in $b_0$.

Only the first term in~\eqref{m1diag} contribute to leading order,
\be
\partial_\sigma\tilde{\bf M}_0=m_1^{\rm LO}b_0\partial_{b_0}\tilde{\bf M}_0 \;,
\ee
where
\be\label{m1LO}
{\bf m}_1^{\rm LO}=\frac{2}{3}\alpha b_0{\bf a}^{\prime2}{\bm1}_4=m_1^{\rm LO}{\bm1}_4 \;.
\ee
The solution is given by the solution to LL, to the power of the moment,
\be
\tilde{\bf M}_0={\bm1}_4\left(\frac{1}{b_0}+\frac{2}{3}\alpha\int_{\sigma}^\infty{\bf a}^{\prime2}\right)^{-m} \;,
\ee
which agrees with~\eqref{Wlead}.

For the next-to-leading order we have
\be\label{M1m0m1m2}
\begin{split}
&\partial_\sigma\tilde{\bf M}_1-m_1^{\rm LO}b_0\partial_{b_0}\tilde{\bf M}_1\\
&=-\left({\bf m}_0-[{\bf m}_1-{\bf m}_1^{\rm LO}]b_0\partial_{b_0}+{\bf m}_2\frac{b_0^2}{2}\partial_{b_0}^2\right)\cdot\tilde{\bf M}_0\\
&=:-\rho\chi_0\left({\bf n}_0+{\bf n}_1b_0\partial_{b_0}+{\bf n}_2\frac{b_0^2}{2}\partial_{b_0}^2\right)\cdot\tilde{\bf M}_0 \;.
\end{split}
\ee
Note that the ${\bf n}_i$ matrices do not depend on $b_0$. 
We can solve~\eqref{M1m0m1m2} 
using the method of characteristics as before. With the notation in~\eqref{Wsumw}, \eqref{characteristicy}, \eqref{wtoomega} and~\eqref{omegam0y}, we find
\be\label{omega1general}
{\bm\omega}_1=\int_\sigma^\infty\!\ud\varphi\bigg\{\frac{{\bf n}_2}{2}\partial_y^2+\frac{{\bf n}_2-{\bf n}_1}{y-J}\partial_y+\frac{{\bf n}_0}{[y-J]^2}\bigg\}\cdot{\bm\omega}_0 \;,
\ee
where ${\bf n}_j$ and $J$ are evaluated at $\varphi$, while $Y=\frac{1}{\rho}+J(\sigma)$.

We can express the non-local integrals above in terms of the following two functions
\be\label{HgenDef}
\mathcal{H}=-\frac{4}{\pi a_0^2}\int\frac{\ud\theta}{\theta\Theta^3}(2+\Delta a^2) 
\ee
and
\be\label{BgenDef}
\mathcal{B}=-\frac{2}{\pi a_0^2}\int\frac{\ud\theta}{\theta\Theta^3}\Delta a^2 \;.
\ee
From~\eqref{m2diag} we find
\be\label{n2gen}
{\bf n}_2=\frac{3\mathcal{H}}{2a_0}{\bf 1}_4 \;.
\ee
From~\eqref{m1diag} and~\eqref{e0m1} we find
\be\label{n1gen}
{\bf n}_1=3{\bf n}_2-\bigg\{\frac{{\bf e}_0}{a_0^3},\frac{3}{2}i{\bm\sigma}_2\cdot\!\left[{\bf a}^{\prime2}{\bf a}'-\frac{{\bf a}'''}{6}\right]+[{\bf a}'\!\cdot\! i{\bm\sigma}_2\!\cdot\!{\bf a}'']\hat{\bf k}\bigg\} \;,
\ee
where $\{{\bf e}_0,{\bf e}_i\}:={\bf e}_0{\bf e}_i+{\bf e}_i{\bf e}_0$. And from~\eqref{m0diag} and~\eqref{e0m0} we find
\be\label{n0gen}
\begin{split}
{\bf n}_0=&-\frac{3\mathcal{B}}{2a_0}({\bf e}_1{\bf e}_1+{\bf e}_2{\bf e}_2)+\frac{3}{2a_0}[\mathcal{H}-2\mathcal{B}]{\bf e}_3{\bf e}_3\\
&+\frac{3}{2a_0^3}{\bf e}_0i{\bm\sigma}_2\cdot\!\left[{\bf a}^{\prime2}{\bf a}'-\frac{{\bf a}'''}{6}\right] \;.
\end{split}
\ee

From ${\bf e}_0\cdot\eqref{omega1general}\cdot{\bf e}_0$ we find
\be\label{e0Me0gen}
\begin{split}
{\bf e}_0\cdot{\bf w}_{m,1}\cdot{\bf e}_0=&\frac{3m\rho}{4a_0[1+\rho J(\sigma)]^{2+m}}\int_\sigma^\infty\ud\sigma' \mathcal{H}(\sigma')\\
\times&\left(1+m+4\frac{1+\rho J(\sigma)}{1+\rho[J(\sigma)-J(\sigma')]}\right) \;.
\end{split}
\ee
Using this to calculate the standard deviation
\be
S^2\approx\chi_0\left[{\bf e}_0\cdot{\bf w}_{2,1}\cdot{\bf e}_0-\frac{2}{1+\rho J}{\bf e}_0\cdot{\bf w}_{1,1}\cdot{\bf e}_0\right] \;,
\ee
gives
\be\label{standardGen}
S^2=\frac{3b_0\rho}{2[1+\rho I(\infty)]^4}\int\ud \sigma \mathcal{H} \;.
\ee
For $a_0\gg1$ we recover~\eqref{standardLCF} as we will now show.

\subsection{LCF}

To check the above results we first consider the LCF regime, $a_0\gg1$. Here we can rescale $\theta=\hat{\theta}/a_0$ and then expand the integrands in~\eqref{HgenDef} and~\eqref{BgenDef} with $\hat{\theta}$ as independent of $a_0$. We find
\be\label{HlcfGen}
\mathcal{H}\approx-\frac{4a_0F^3}{\pi}\int_{-\infty}^\infty\ud\theta\frac{2+\theta^2}{\theta^4\left(1+\frac{\theta^2}{12}\right)^3}=\frac{55a_0F^3}{24\sqrt{3}} \;,
\ee
where the integration contour goes above the pole at $\theta=0$ but below $\theta=i\sqrt{12}$,
and similarly
\be\label{BlcfGen}
\mathcal{B}\approx\frac{5\sqrt{3}}{8}a_0F^3 \;.
\ee
Restricting for simplicity to linear polarization and spin parallel (or antiparallel) to the magnetic field, we find
\be\label{n0n1LCF}
{\bf n}_0=F^3
\begin{pmatrix}
0&\frac{3}{2}\epsilon\\0&-\frac{15\sqrt{3}}{16} 
\end{pmatrix} 
\qquad
{\bf n}_1=F^3
\begin{pmatrix}
\frac{55\sqrt{3}}{16}&-\frac{3}{2}\epsilon\\-\frac{3}{2}\epsilon&\frac{55\sqrt{3}}{16} 
\end{pmatrix}
\ee
and
\be
{\bf n}_2=\frac{55}{16\sqrt{3}}F^3\begin{pmatrix}1&0\\0&1\end{pmatrix} \;,
\ee
with the same notation as in Sec.~\ref{LCF of moments}. By plugging these results into~\eqref{omega1general} we recover~\eqref{omegam1}.

\subsection{Perturbative limit}

For $a_0\ll1$ we expand $\mathcal{H}$ and $\mathcal{B}$ to $\mathcal{O}(a_0^2)$. A $\mathcal{O}(1/a_0^2)$ term in $\mathcal{H}$ vanishes upon integrating over $\theta$. We Fourier transform the field,
\be
{\bf f}(\phi)=\int\frac{\ud w}{2\pi}{\bf f}(w)e^{-iw\phi} \;,
\ee
where ${\bf f}(\phi)={\bf a}(\phi)/a_0$. All terms are quadratic in the field, so we have two Fourier variables, $w_1$ and $w_2$, but the $\sigma$ integral gives a delta function $\delta(w_1+w_2)$. The $\theta$ integral can, after partial integration, be performed using 
\be
\int\frac{\ud\theta}{\theta}\sin(w\theta)=\pi\text{sign}(w) \;.
\ee 
We find
\be\label{HsigmaIntFourier}
\int\ud \sigma\{\mathcal{H},\mathcal{B}\}=\left\{\frac{28}{15},\frac{4}{3}\right\}\int_0^\infty\frac{\ud w}{2\pi}w^3{\bf f}(w)\cdot{\bf f}(-w) \;.
\ee

If we expand~\eqref{e0Me0gen} to $\mathcal{O}(\alpha)$, set $m=1$ and use~\eqref{HsigmaIntFourier} then we find agreement with what one finds by expanding Eq.~(3.21) in~\cite{Gonzo:2023cnv} to $\mathcal{O}(\hbar)$.

If the field components are proportional to $\sin(\phi)$ or $\cos(\phi)$ times a slowly varying envelope, then ${\bf f}(w)$ is sharply peaked at $w=\pm1$ and we can approximate $w^3\to1$ and transform back from $w$ to $\sigma$
\be\label{HBperGen}
\int\ud \sigma\{\mathcal{H},\mathcal{B}\}=\int\ud\sigma\left\{\frac{14}{15},\frac{2}{3}\right\}\frac{{\bf a}^2(\sigma)}{a_0^2} \;.
\ee

\subsection{LMF circular polarization}

Next we consider the locally-monochromatic-field (LMF) regime for a circularly polarized field,
\be\label{ahCirc}
{\bf a}(\phi)=a_0 h\left(\frac{\phi}{\mathcal{T}}\right)(\sin(\phi){\bf e}_1+\cos(\phi){\bf e}_2) \;,
\ee
where $h$ is an envelope function, e.g. $h(x)=e^{-x^2}$, and $\mathcal{T}\gg1$. We do not assume that $a_0$ is large. By rescaling $\sigma=\mathcal{T}u$ we can expand the integrand to leading order in $1/\mathcal{T}\ll1$ with $u$ and $\theta$ as independent of $\mathcal{T}$~\cite{Torgrimsson:2020gws}. For example, this gives
\be
J(u)=\mathcal{T}\int_u^{\infty}\ud v\, h^2(v) \;.
\ee
When convenient we denote $a(u)=a_0h(u)$. We consider here for simplicity spin parallel to $\hat{\bf k}$, because spin components in the ${\bf e}_1$-${\bf e}_2$ plane rotate. 

We begin with the off-diagonal, local terms. The ones from ${\bf n}_0$ in~\eqref{n0gen} average out. From~\eqref{n1gen} we find to leading order in $\mathcal{T}\gg1$,
\be\label{n1offCirc}
{\bf n}_1=\frac{h^2(u)}{a_0}\{{\bf e}_0,\hat{\bf k}\}+\text{diagonal terms} \;.
\ee
Inserting this into~\eqref{omega1general} gives the off diagonal terms
\be\label{e0M1e3circ}
\begin{split}
{\bf w}_1|_\text{off}
&=\frac{m\{{\bf e}_0,\hat{\bf k}\}}{(\rho y)^{m+1}a_0}\int_u^\infty\ud v\frac{\ud}{\ud v}\ln[y-J(v)]\\
&=\frac{m}{a_0}\frac{\ln[1+\rho J(u)]}{[1+\rho J(u)]^{1+m}}\{{\bf e}_0,\hat{\bf k}\} \;.
\end{split}
\ee
For a constant envelope and $m=1$ we recover Eq.~(39) in~\cite{Torgrimsson:2021zob}. To compare with an experiment we would only need~\eqref{e0M1e3circ} evaluated at $u=-\infty$, but the above expression for finite $u$ can be used to compare with a numerical solution of~\eqref{MnEq}. \eqref{e0M1e3circ} vanishes for $m=0$, so we will treat this case separately in the next subsection by going to one order higher in $b_0$. 

Now we turn to the diagonal, nonlocal terms. Expanding to leading order in $1/\mathcal{T}$ gives
\be
\Delta a^2=4a^2(u)\sin^2\frac{\theta}{2} 
\ee
and
\be\label{M2circ}
M^2=1+a^2(u)\left(1-\text{sinc}^2\frac{\theta}{2}\right) \;.
\ee
In order to facilitate comparison with the monochromatic ($h$ constant) results in~\cite{Torgrimsson:2021zob}, we write
\be
\mathcal{H}=\mathcal{I}_1-\mathcal{I}_0
\qquad
\mathcal{B}=\frac{1}{2}\mathcal{I}_1 \;,
\ee 
where
\be\label{I0def}
\mathcal{I}_0=\frac{1}{a_0^2}\int_0^\infty\ud\gamma\,\gamma^2\mathcal{J}_0
=\frac{8}{\pi a_0^2}\int\frac{\ud\theta}{\theta\Theta^3}
\ee
and
\be\label{I1def}
\mathcal{I}_1=\frac{1}{a_0^2}\int_0^\infty\ud\gamma\,\gamma^2\mathcal{J}_1
=-\frac{16h^2}{\pi}\int\frac{\ud\theta}{\theta\Theta^3}\sin^2\frac{\theta}{2} \;.
\ee
We have presented the expressions with $\mathcal{J}_n$ to compare with~\cite{Torgrimsson:2021zob}, but here we will only work with the $\theta$ integrals. $\mathcal{I}_0$ and $\mathcal{I}_1$ have a nontrivial dependence on $a_0$.


\begin{figure}
\includegraphics[width=\linewidth]{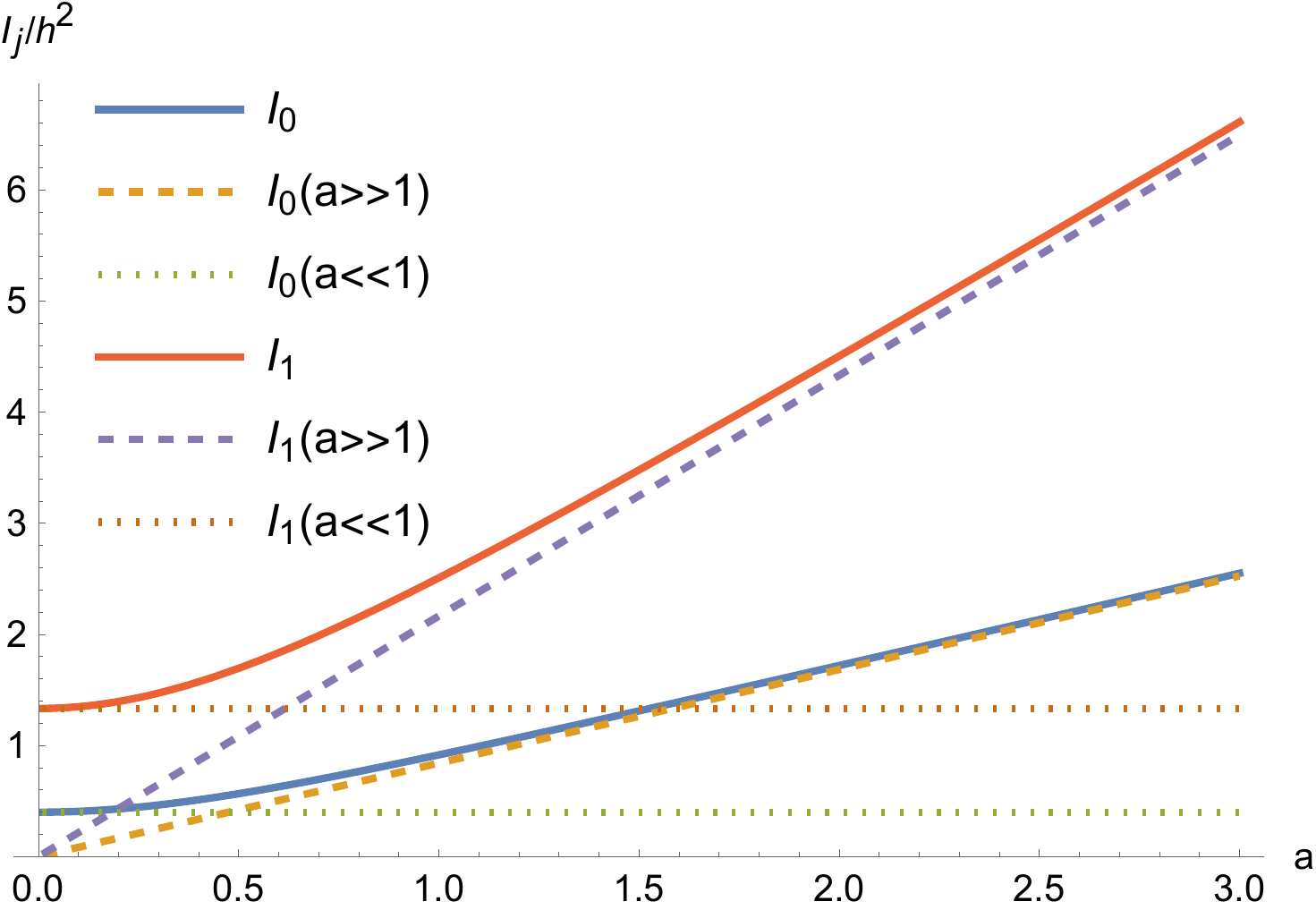}
\caption{Solid lines have been obtained by numerically integrating~\eqref{I0def} and~\eqref{I1def} (along a line parallel to the real axis above the pole at $\theta=0$ but below the additional poles due to $\Theta$), dashed lines show~\eqref{I0I1LCF}, and dotted lines show~\eqref{I0I1per}.}
\label{I0I1fig}
\end{figure}

For $a_0\gg1$ we find
\be\label{I0I1LCF}
\mathcal{I}_0\approx\frac{35}{24\sqrt{3}}a_0h^3(u)
\qquad
\mathcal{I}_1\approx\frac{5\sqrt{3}}{4}a_0h^3(u) \;.
\ee 
\eqref{I0I1LCF} agrees with~\eqref{HlcfGen} and~\eqref{BlcfGen}.
For $a_0\ll1$ we can expand the integrand without rescaling $\theta$ and then perform partial integration until we have $\int\sin(\theta)/\theta=\pi$, which gives
\be\label{I0I1per}
\mathcal{I}_0\approx\frac{2}{5}h^2(u)
\qquad
\mathcal{I}_1\approx\frac{4}{3}h^2(u) \;.
\ee
\eqref{I0I1per} agrees with~\eqref{HBperGen}.
These $a_0\gg1$ and $a_0\ll1$ approximations are compared with the exact result in Fig.~\ref{I0I1fig}.

From~\eqref{e0Me0gen} we find
\be\label{e0Me0circ}
\begin{split}
{\bf e}_0\cdot{\bf w}_1\cdot{\bf e}_0=&\frac{3m\rho}{4a_0[1+\rho J(u)]^{2+m}}\mathcal{T}\int_u^\infty\ud v(\mathcal{I}_1-\mathcal{I}_0)(v)\\
\times&\left(1+m+4\frac{1+\rho J(u)}{1+\rho[J(u)-J(v)]}\right)
\end{split}
\ee
and from the ${\bf e}_3{\bf e}_3$ term in~\eqref{n0gen} we find
\be
\begin{split}
&\hat{\bf k}\cdot{\bf w}_1\cdot\hat{\bf k}={\bf e}_0\cdot{\bf w}_1\cdot{\bf e}_0\\
&-\frac{3\rho}{2a_0[1+\rho J(u)]^m}\mathcal{T}\int_u^\infty\ud v\frac{\mathcal{I}_0(v)}{(1+\rho[J(u)-J(v)])^2} \;.
\end{split}
\ee

\begin{figure}
\includegraphics[width=\linewidth]{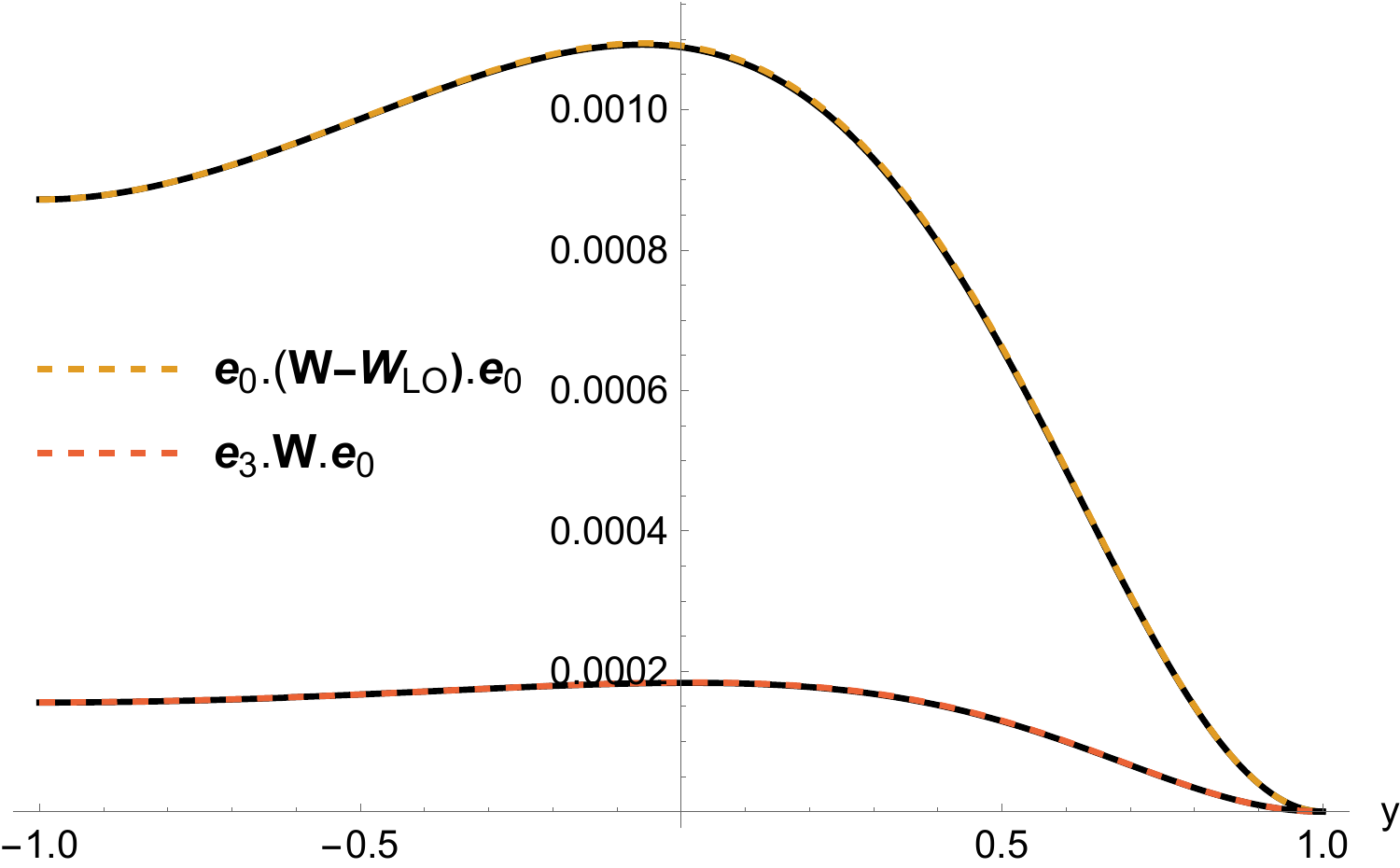}
\caption{Comparison of the numerical solution of~\eqref{Weq} and the low-energy approximations in~\eqref{e0M1e3circ} and~\eqref{e0Me0circ}, for $m=1$ and a circularly polarized field with a Sauter pulse envelope, with $a_0=1$, $b_0=10^{-3}$ and $\rho\mathcal{T}=1$.}
\label{circFig}
\end{figure}

For $m=0$ and $u\to-\infty$ this reduces to
\be\label{kMkCirc}
\hat{\bf k}\cdot\chi_0{\bf w}_1\cdot\hat{\bf k}=-\frac{3}{2}b_0\rho\mathcal{T}\int\ud v\frac{\mathcal{I}_0}{[1+\rho I(v)]^2} \;,
\ee
where
\be
I(u)=\mathcal{T}\int_{-\infty}^u\ud v\, h^2(v) \;.
\ee
For a constant envelope we recover Eq.~(47) in~\cite{Torgrimsson:2021zob}. Fig.~\ref{circFig} shows that the above low-energy approximations agree well with a full numerical solution of~\eqref{Weq} for a Sauter-pulse envelope, $h(u)=\text{sech}^2(u)$.

For $a_0\sim1$ we have $\rho\ll1$, but, as we can see explicitly in the above expressions, the relevant parameter in this regime is $\rho\mathcal{T}$, which can be $\mathcal{O}(1)$ for sufficiently long pulse, $\mathcal{T}\gg1$, which means we still need to resum the $\alpha$ expansion.

The standard deviation is obtained directly from~\eqref{standardGen}.
Using~\eqref{I0I1LCF}, we find for $a_0\gg1$
\be
S^2=\frac{55\chi_0\rho}{16\sqrt{3}[1+\rho I(\infty)]^4}\mathcal{T}\int\ud u\,h^3(u) \;,
\ee
which agrees with what one finds by taking the LCF limit first~\eqref{standardLCF}. For $a_0\ll1$ we find, using~\eqref{I0I1per},
\be\label{StandardLMFcircPer}
S^2=\frac{7b_0\rho}{5[1+\rho I(\infty)]^4}\mathcal{T}\int\ud u\, h^2(u) \;.
\ee

\subsection{LMF -- off-diagonal element for $m=0$}

The zeroth moment is special because the zeroth order in $\chi_0$ is just the identity matrix,
\be
\tilde{\bf M}={\bf1}+\Delta{\bf M} \;.
\ee
Because of unitarity we also have $\Delta{\bf M}\cdot\{1,{\bf 0}\}={\bf 0}$, since if we sum the zeroth moment over the final spin then we have summed the probability over all possible final states. Thus, if we restrict to spin parallel to the laser wave vector $\hat{\bf k}$, then a general Mueller matrix would be $2\times2$, but, for the zeroth moment, two of the four elements are trivial
\be
\Delta{\bf M}=\begin{pmatrix}0& B\\0& A\end{pmatrix} \;.
\ee
$A$ is given by~\eqref{kMkCirc} and is $\mathcal{O}(b_0)$. \eqref{e0M1e3circ} shows that $B$ vanishes at $\mathcal{O}(b_0)$. We will now show that $B=\mathcal{O}(b_0^2)$. 

From~\eqref{MnEq} we have
\be
\begin{split}
&\partial_\sigma B=\\
&-\int_0^1\ud q\bigg[{\bf e}_0\!\cdot\!{\bf M}_L\!\cdot\!{\bf e}_0 B(b_0)+{\bf e}_0\!\cdot\!{\bf M}_C\!\cdot\!{\bf e}_0B([1-q]b_0)\bigg] \\
&-\int_0^1\ud q\bigg[{\bf e}_0\!\cdot\!{\bf M}_L\!\cdot\!{\bf e}_3 A(b_0)+{\bf e}_0\!\cdot\!{\bf M}_C\!\cdot\!{\bf e}_3A([1-q]b_0)\bigg] \;.
\end{split}
\ee
Since ${\bf e}_0\!\cdot\!{\bf M}_L\!\cdot\!{\bf e}_0=-{\bf e}_0\!\cdot\!{\bf M}_C\!\cdot\!{\bf e}_0$ and ${\bf e}_0\!\cdot\!{\bf M}_L\!\cdot\!{\bf e}_3=-{\bf e}_0\!\cdot\!{\bf M}_C\!\cdot\!{\bf e}_3$ we have to leading order
\be
\partial_\sigma B=f_1b_0\partial_{b_0}B+f_2b_0\partial_{b_0}A \;,
\ee
where, from~\eqref{m1LO} and~\eqref{n1offCirc},
\be
\begin{split}
f_1&=\int\ud q\, q\,{\bf e}_0\!\cdot\!{\bf M}_C\!\cdot\!{\bf e}_0\approx\frac{2}{3}\alpha b_0{\bf a}^{\prime2}(\sigma)\approx\frac{2}{3}\alpha b_0a^2(u)\\
f_2&=\int\ud q\, q\,{\bf e}_0\!\cdot\!{\bf M}_C\!\cdot\!{\bf e}_3\approx-\frac{2}{3}\alpha b_0^2 a^2(u) \;.
\end{split}
\ee
Using the method of characteristics as above, we find
\be
B=-3b_0^2\rho\mathcal{T}\int_u^\infty\ud v\,\mathcal{I}_0(v)\frac{\ln(1+\rho[J(u)-J(v)])}{(1+\rho[J(u)-J(v)])^3} \;.
\ee
For a constant envelope $h=1$, we recover\footnote{$\mathcal{T}$ in~\cite{Torgrimsson:2021zob} is equal to $\alpha\mathcal{T}$ in the notation here.} Eq.~(48) in~\cite{Torgrimsson:2021zob}.
If the initial state is unpolarized, then $B(u\to-\infty)$ gives the degree of longitudinal polarization in the final state. It is small due to the factor of $b_0^2$, but it shows that for circular polarization  there is a nonzero induced (longitudinal) polarization, in contrast to linear polarization where the induced polarization averages out due to the oscillations of the field.

\subsection{LMF linear polarization}

In this this section we consider fields with linear polarization which oscillate in a symmetric way, so that some spin effects average out.
The spin components in the $\hat{\bf E}$-$\hat{\bf k}$ plane rotates.
There is no induced polarization (Sokolov-Ternov effect) in this case since the components, which would otherwise lead to such effects in the $\hat{\bf B}$ direction, oscillate and average out. However, this does not mean that there cannot be any nontrivial change in the spin along $\hat{\bf B}$. If the initial state is polarized then there can be depolarization, since the diagonal element $\hat{\bf B}_0\cdot\tilde{\bf M}\cdot\hat{\bf B}_0$ is different from ${\bf e}_0\cdot\tilde{\bf M}\cdot{\bf e}_0$. From the ${\bf e}_2{\bf e}_2$ term in~\eqref{n0gen} we find
\be\label{B0M1B0B}
\begin{split}
&\hat{\bf B}_0\cdot{\bf w}_1\cdot\hat{\bf B}_0={\bf e}_0\cdot{\bf w}_1\cdot{\bf e}_0\\
&-\frac{3\rho}{2a_0[1+\rho J(\sigma)]^m}\int_\sigma^\infty\ud\sigma'\frac{\mathcal{B}(\sigma')}{(1+\rho[J(\sigma)-J(\sigma')])^2} \;,
\end{split}
\ee
where ${\bf e}_0\cdot{\bf w}_1\cdot{\bf e}_0$ is given by~\eqref{e0Me0gen}.
It is not obvious at this point, but we will show below that $\mathcal{B}>0$, so for $m=0$, the ${\bf e}_0$ and $\hat{\bf B}_0$ components of the Mueller matrix are given by
\be
\tilde{\bf M}_0+\tilde{\bf M}_1=\begin{pmatrix}1&0\\0&1-\delta\end{pmatrix} \;,
\ee
where $\delta>0$. If the initial state has ${\bf N}_0=\{1,p\}$ with $0\leq|p|\leq1$, then ${\bf N}_0\cdot\tilde{\bf M}\approx\{1,p(1-\delta)\}$, so the final state is less polarized. 

Now we consider a field on the form
\be\label{ahLin}
{\bf a}(\phi)=a_0h\left(\frac{\phi}{\mathcal{T}}\right)\sin(\phi+\Delta){\bf e}_1 \;,
\ee
where $\Delta$ is a constant and $h$ is an envelope similar to the circular case~\eqref{ahCirc}. We can again find the leading order in $1/\mathcal{T}\ll1$ by rescaling $\sigma=\mathcal{T}u$ and consider $u$ and $\theta$ as independent of $\mathcal{T}$. We have
\be\label{JuLinMono}
J(u)\approx\mathcal{T}\int_u^\infty\!\ud v[h(v)\cos(\mathcal{T}v+\Delta)]^2\approx\frac{\mathcal{T}}{2}\int_u^\infty\!\ud v\, h^2(v) \;.
\ee
For the effective mass one finds
\be\label{M2lin}
\begin{split}
M^2=&1+\frac{a^2(u)}{2}\bigg(1-\text{sinc}(\theta)\\
&+2\sin^2(\mathcal{T}u+\Delta)\left[\text{sinc}(\theta)-\text{sinc}^2\left(\frac{\theta}{2}\right)\right]\bigg) \;.
\end{split}
\ee
$M^2$ only depends on $u$ via the envelope $a(u)$ in the circular case~\eqref{M2circ}, but~\eqref{M2lin} also has a rapidly oscillating term. If the $\sigma$ integrand had been linear in $M^2$ we could have replaced $\sin^2(\mathcal{T}u+\Delta)\to1/2$, and then~\eqref{M2lin} would have reduced to the circular case~\eqref{M2circ} but with $a_0\to a_0/\sqrt{2}$. However, the integrands in~\eqref{HgenDef} and~\eqref{BgenDef} are not linear in $M^2$. Instead we have integrands on the form
\be
F[a(u),\sin^2(\mathcal{T}u+\Delta)] \;,
\ee     
where $F$ is a nonlinear function. We introduce a temporary variable $\delta$ and then Taylor expand
\be\label{Ftaylor}
F[a,\delta\sin^2(\mathcal{T}u+\Delta)]=\sum_{n=0}^\infty F_n[a]\delta^n\sin^{2n}(\mathcal{T}u+\Delta) \;.
\ee
Next we write
\be\label{binomial}
\begin{split}
\sin^{2n}(\varphi)&=\left(\frac{e^{i\varphi}-e^{-i\varphi}}{2i}\right)^{2n}\\
&=\frac{(2n)!}{4^n(n!)^2}+\text{oscillating terms} \;.
\end{split}
\ee
We can neglect the oscillating terms since they oscillate rapidly for $\mathcal{T}\gg1$. The sum  
\be
\sum_{n=0}^\infty F_n[a]\delta^n\frac{(2n)!}{4^n(n!)^2}
\ee
might look more complicated than~\eqref{Ftaylor}, but Mathematica has no problem summing this series in the cases we have considered. Applying this to~\eqref{BgenDef} and~\eqref{HgenDef} gives
\be\label{BlinGen}
\mathcal{B}=-\frac{h^2(u)}{\pi}\int\ud\theta\frac{(4X+3Y)\sin^2\left(\frac{\theta}{2}\right)}{X^{5/2}(X+Y)^{3/2}\theta^4} 
\ee
and
\be\label{HlinGen}
\mathcal{H}=2\mathcal{B}-\frac{1}{\pi a_0^2}\int\ud\theta\frac{8X^2+8XY+3Y^2}{\theta^4X^{5/2}(X+Y)^{5/2}} \;,
\ee
where
\be
\begin{split}
X&=1+\frac{a^2(u)}{2}[1-\text{sinc}(\theta)]\\
Y&=a^2(u)\left[\text{sinc}(\theta)-\text{sinc}^2\left(\frac{\theta}{2}\right)\right] \;.
\end{split}
\ee
We can obtain the same results by replacing
\be\label{integralOnePeriod}
F[a(u),\sin^2(\mathcal{T}u+\Delta)]\to\frac{1}{\pi}\int_{-\pi/2}^{\pi/2}\ud\varphi\, F[a(u),\sin^2(\varphi)] \;,
\ee
since the integral selects the terms in~\eqref{binomial} which do not oscillate. The approach using~\eqref{integralOnePeriod} can be found in the literature~\cite{Dinu:2013hsd}.
Note that $\mathcal{B}/h^2$ and $\mathcal{H}/h^2$ are only functions of one variable, $a(u)$, so one can make numerical interpolation functions of them for $0<a<a_0$ without choosing any specific envelope $h(u)$, and then afterwards one can apply the interpolation to e.g. $h=e^{-u^2}$, $h=\text{sech}^2(u)$ or any other envelope.  

For $a_0\gg1$ we can perform the $\theta$ integral by rescaling $\theta\to\theta/a$ and then expanding the integrand to leading order,
\be\label{BLCF}
\mathcal{B}\approx-\frac{6\sqrt{3}}{\pi}a_0h^3\int\ud\theta\frac{48+\theta^2}{\theta^2(12+\theta^2)^{5/2}}=\frac{5a_0h^3}{2\sqrt{3}\pi} 
\ee
and similarly
\be\label{HLCF}
\mathcal{H}\approx\frac{55a_0h^3}{18\sqrt{3}\pi} \;.
\ee
Plugging~\eqref{BLCF} into~\eqref{B0M1B0B} gives for $m=0$
\be\label{B0M1B0linLCF}
\hat{\bf B}_0\cdot\chi_0{\bf w}_1\cdot\hat{\bf B}_0=-\frac{5\sqrt{3}\chi_0\rho\mathcal{T}}{4\pi}\int\ud u\frac{h^3(u)}{[1+\rho I(u)]^2} \;,
\ee
where
\be
I(u)=\frac{\mathcal{T}}{2}\int_{-\infty}^u\ud v\, h^2(v) \;.
\ee
Plugging~\eqref{HLCF} into~\eqref{standardGen} gives
\be\label{standardLinLCF}
S^2=\frac{55\chi_0\rho\mathcal{T}}{12\sqrt{3}\pi[1+\rho I(\infty)]^4}\int\ud u\, h^3(u) \;.
\ee

\eqref{B0M1B0linLCF} agrees with the diagonal component of~\eqref{w01}, and~\eqref{standardLinLCF} agrees with~\eqref{standardLCF}, but to see that we need to simplify~\eqref{w01} and~\eqref{standardLCF}. We have $F(\sigma)\approx h(u)|\cos(\mathcal{T}u+\Delta)|$, which makes the integral in~\eqref{w01} and~\eqref{standardLCF} non-analytic. We can write them as
\be\label{absCos3int}
\begin{split}
&\int\ud\sigma h^3\left(\frac{\sigma-\Delta}{\mathcal{T}}\right)|\cos^3(\sigma)|G\left(\frac{\sigma-\Delta}{\mathcal{T}}\right)\\
&\approx\sum_{n=-\infty}^\infty h^3\left(\frac{n\pi}{\mathcal{T}}\right)G\left(\frac{n\pi}{\mathcal{T}}\right)\int_{-\pi/2}^{\pi/2}\ud\sigma\cos^3(\sigma)\\
&\approx\frac{4\mathcal{T}}{3\pi}\int\ud u\, h^3(u)G(u) \;,
\end{split}
\ee
where $G$ just stands for the rest of the integrand. Thus, we should multiply~\eqref{HlcfGen} and~\eqref{BlcfGen} by $4/(3\pi)$ and replace $F\to h$ and then we find agreement with~\eqref{HLCF} and~\eqref{BLCF}.
And by using~\eqref{absCos3int} on~\eqref{w01} and~\eqref{standardLCF} we recover~\eqref{B0M1B0linLCF} and~\eqref{standardLinLCF}. 

\begin{figure}
\includegraphics[width=\linewidth]{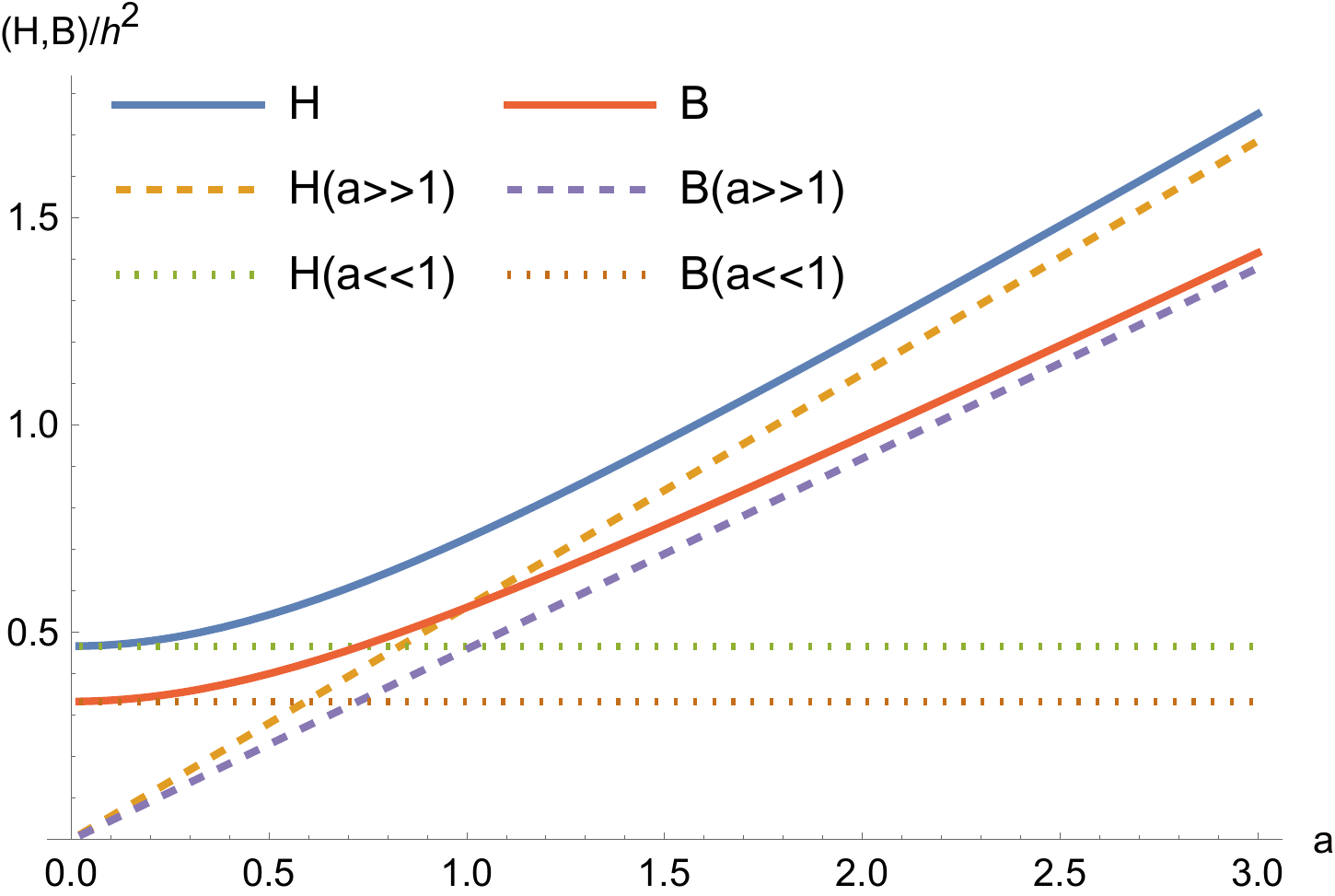}
\caption{$\mathcal{B}$ and $\mathcal{H}$ divided by $h^2$. Solid lines are obtained by numerically integrating~\eqref{BlinGen} and~\eqref{HlinGen}, dashed lines show~\eqref{BLCF} and~\eqref{HLCF}, and dotted lines show~\eqref{HBper}.}
\label{HBfig}
\end{figure}

For $a_0\ll1$ we can expand the integrands in~\eqref{BlinGen} and~\eqref{HlinGen} without rescaling $\theta$. We find
\be\label{HBper}
\mathcal{B}=\frac{h^2}{3}
\qquad
\mathcal{H}=\frac{7h^2}{15} \;.
\ee
\eqref{HBper} agrees with~\eqref{HBperGen}.
The perturbative~\eqref{HBper} and LCF approximations~\eqref{BLCF} and~\eqref{HLCF} are compared with the exact result in Fig.~\ref{HBfig}.

\subsection{Bichromatic field}

For a field which is linearly polarized and almost monochromatic, as in~\eqref{ahLin}, the off-diagonal terms in~\eqref{n1gen} and~\eqref{n0gen} tend to average out due to the oscillations, which means there is no (or a negligible) induced polarization. However, this is not necessarily the case if we instead consider an almost bichromatic field
\be
{\bf a}(\phi)=a_0h\left(\frac{\phi}{\mathcal{T}}\right)[a_1\sin(\phi)+a_2\sin(2\phi+\Delta)]{\bf e}_1 \;.
\ee 
Such fields were studied in~\cite{Seipt:2019ddd,Chen:2019vly}, for the same reason. 
Instead of~\eqref{JuLinMono} we have
\be
J\approx(a_1^2+4a_2^2)\frac{\mathcal{T}}{2}\int_u^\infty\ud v\, h^2(v) \;.
\ee
The off-diagonal terms in~\eqref{n1gen} and~\eqref{n0gen} are proportional to
\be\label{acube}
\begin{split}
&\frac{a^{\prime3}(\sigma)}{(a_0h)^3}\approx a_1^3\cos^3(\sigma)+6a_1^2a_2\cos^2(\sigma)\cos(2\sigma+\Delta)\\
&+12a_1a_2^2\cos(\sigma)\cos^2(2\sigma+\Delta)+8a_2^3\cos^3(2\sigma+\Delta) \;.
\end{split}
\ee
After writing $\cos$ in terms of $e^{...}$, one finds that only the second term contains a non-oscillating part, 
\be
\eqref{acube}\approx\frac{3}{2}a_1^2a_2\cos\Delta+\text{oscillating terms} \;.
\ee
So, from~\eqref{n1gen} and~\eqref{n0gen} we find 
\be
{\bf n}_1\approx\frac{9}{4}a_1^2a_2\cos(\Delta)h^3\{{\bf e}_0,{\bf e}_2\}+\text{diagonal terms}
\ee
and
\be
{\bf n}_0\approx-\frac{9}{4}a_1^2a_2\cos(\Delta)h^3{\bf e}_0{\bf e}_2+\text{diagonal terms} \;.
\ee
Since these off-diagonal terms are already local, they agree\footnote{Note ${\bf e}_2=-\hat{\bf B}_0$ as explained in~\cite{Torgrimsson:2020gws}.} with what one finds from the off-diagonal terms of the LCF approximation in~\eqref{n0n1LCF} even without making any expansion in $1/a_0$. For $m=0$ we find
\be\label{inducedPolBi}
{\bf e}_0\cdot{\bf w}_1\cdot{\bf e}_2=-\frac{9}{4}a_1^2a_2\rho\cos\Delta\int\ud u\frac{h^3(u)}{[1+\rho I(u)]^2} \;,
\ee
where
\be
I(u)=(a_1^2+4a_2^2)\frac{\mathcal{T}}{2}\int_{-\infty}^u\ud v\, h^2(v) \;.
\ee
\eqref{inducedPolBi} gives the polarization of the final state if the initial state is unpolarized.

\section{Spin perpendicular to magnetic field}\label{Spin perpendicular to magnetic field}

In this section we consider the components of the Stokes vector which describe spin perpendicular to the magnetic field. We consider a linearly polarized field, which means the ${\bf e}_0$ and $\hat{\bf B}_0$ components, which we studied in Sec.~\ref{LCF of moments}, decouple from the $\hat{\bf E}_0$ and $\hat{\bf k}$ components, which we will now study. Here $\hat{\bf E}_0$ is the constant axis parallel to the electric field, $\hat{\bf k}$ is the axis of propagation, and $\hat{\bf B}_0=\hat{\bf E}_0\times\hat{\bf k}$. The off-diagonal elements of ${\bf M}_L$ in the $\hat{\bf E}_0$-$\hat{\bf k}$ space act as a rotation matrix,
\be
{\bf R}(\sigma,b_0)=\frac{\alpha}{b_0}(\hat{\bf k}\hat{\bf E}-\hat{\bf E}\hat{\bf k})\int_0^1\ud q\,q\frac{\text{Gi}(\xi)}{\sqrt{\xi}} \;,
\ee
where $\xi=(r/\chi(\sigma))^{2/3}$, $r=(1/s)-1$, $s=1-q$ and $\text{Gi}$ is the Scorer function\footnote{$\text{Gi}$ also appears in the results in~\cite{Ilderton:2020gno} for the loop, though not as the cross term between the loop and the zeroth order and hence not as a Mueller matrix.}. For linear polarization we have
\be
\hat{\bf k}\hat{\bf E}-\hat{\bf E}\hat{\bf k}=\epsilon(\sigma)(\hat{\bf k}\hat{\bf E}_0-\hat{\bf E}_0\hat{\bf k}) \;,
\ee
so ${\bf R}$ is proportional to a constant matrix. So, if we write
\be
{\bf M}=\exp\left\{\int_\sigma^\infty\ud\sigma'{\bf R}(\sigma')\right\}{\bf M}'
\ee
then
\be
\frac{\partial{\bf M}}{\partial\sigma}+{\bf R}\cdot{\bf M}=\exp\left\{\int_\sigma^\infty\ud\sigma'{\bf R}(\sigma')\right\}\frac{\partial{\bf M}'}{\partial\sigma} \;,
\ee
which, at first sight, might suggest that by multiplying~\eqref{MnEq} with $e^{-\int...}$ one would obtain an equation for ${\bf M}'$ which does not involve ${\bf R}$. However, that is not the case, because in the term with ${\bf M}_C$ we would have
\be
\begin{split}
&\exp\left\{-\int_\sigma^\infty\ud\sigma'{\bf R}(\sigma',b_0)\right\}\\
&\cdot\exp\left\{-\int_\sigma^\infty\ud\sigma'{\bf R}(\sigma',[1-q]b_0)\right\}\ne1 \;.
\end{split}
\ee
To resolve this issue we consider the $\chi\ll1$ expansion of
\be\label{GintGi}
\mathcal{G}(\chi):=\int_0^1\ud q\,q\frac{\text{Gi}(\xi)}{\sqrt{\xi}} \;.
\ee
To find a $\chi$ expansion we calculate the Mellin transform\footnote{Mellin transforms have also been used in e.g.~\cite{BaierMellin,RitusMellin}.}
\be
\mathcal{G}(S)=\int_0^\infty\ud\chi\,\chi^{S-1}\mathcal{G}(\chi) \;.
\ee
We change variable from $q$ to $\xi$, so that $\chi$ does not appear in $\text{Gi}$, and then we perform the $\chi$ integral first. We find
\be
\mathcal{G}(S)=\int_0^\infty\ud\xi\frac{3\pi}{4}\frac{S(1+S)}{\sin(\pi S)}\xi^{-\frac{3}{2}(1+S)}\text{Gi}(\xi)
\ee
provided $-2<\text{Re }S<1$. Next we express the Scorer function as an integral,
\be
\text{Gi}(\xi)=\int_0^\infty\frac{\ud\tau}{\pi}\sin\left(\frac{\tau^3}{3}+\xi\tau\right) \;.
\ee
Then we perform first the $\xi$ integral and then the $\tau$ integral. For
\be\label{ReS}
-1<\text{Re }S<-\frac{1}{3}
\ee
we find
\be\label{MellinGgrand}
\mathcal{G}(S)=\frac{\pi}{4}3^{(1+S)/2}S\frac{\Gamma\left[-\frac{1+3S}{2}\right]}{\Gamma\left[-\frac{1+S}{2}\right]\cos^2\left[\frac{\pi}{2}S\right]} \;.
\ee
The inverse Mellin transform is given by
\be\label{MellinGinverse}
\mathcal{G}(\chi)=\int\frac{\ud S}{2\pi i}\chi^{-S}\mathcal{G}(S) \;,
\ee
where the integration contour crosses the real axis at some point in the range~\eqref{ReS}. If we close the contour around the real axis with $S>-1/3$ we obtain a $\chi\gg1$ expansion. Here we are instead interested in the $\chi\ll1$ expansion, which we obtain by closing the contour around the real axis with $S<-1$. We have a simple pole at $S=-1$, which gives the leading order, and double poles at $S=-3,-5,-7,\dots$, which give higher orders. We find~\cite{Torgrimsson:2020gws} 
\be
\mathcal{G}(\chi)=\mathcal{G}_{LO}(\chi)+\mathcal{O}(\chi^3\ln(\chi)) \;,
\ee  
where the leading order is (cf. e.g.~\cite{BaierSokolovTernov})
\be
\mathcal{G}_{LO}(\chi)=\frac{\chi}{2\pi} \;.
\ee
The higher orders are given by
\be
\begin{split}
&\Delta\mathcal{G}(\chi)=\sum_{n=1}^\infty\frac{\sqrt{3}}{2\pi}\left(\frac{\chi}{\sqrt{3}}\right)^{2n+1}\frac{\Gamma[1+3n]}{\Gamma[n]}(1+2n)\\
&\times\bigg(2\ln\left(\frac{\chi}{\sqrt{3}}\right)-2\gamma_E+\frac{2}{1+2n}+3H_{3n}-H_{n-1}\bigg) \;,
\end{split}
\ee
where $\Delta\mathcal{G}=\mathcal{G}-\mathcal{G}_{LO}$, $\gamma_E=0.577...$ is Euler's constant and $H_n=\sum_{k=1}^n 1/k$ is the harmonic number. The coefficients $a_n$ multiplying $(\chi/\sqrt{3})^{2n+1}$ grow asymptotically as $a_n/a_{n+1}\sim1/(27n^2)$, so this is an asymptotic series. This series has a different mathematical structure compared to the corresponding series for the ${\bf e}_0$ and $\hat{\bf B}_0$ components, which e.g. do not have $\ln(\chi)$ terms. 

A convenient and fast way to numerically compute $\mathcal{G}(\chi)$ exactly, is to integrate~\eqref{MellinGinverse} along a contour $C_1$ which is parallel to the imaginary axis with e.g. $\text{Re}(S)=-1/2$. In our numerical implementation (with Mathematica), this is in fact much faster than directly integrating~\eqref{GintGi}. 
An equally convenient way to compute $\Delta\mathcal{G}(\chi)$ is to integrate~\eqref{MellinGinverse} with exactly the same integrand~\eqref{MellinGgrand} but along a contour $C_2$ which is parallel to the imaginary axis with $\text{Re}(S)=-2$. This works because we can deform $C_1$ to $C_2$ plus a small loop around the pole at $S=-1$, and the loop integral gives exactly $\mathcal{G}_{LO}$, so omitting this loop gives $\mathcal{G}-\mathcal{G}_{LO}$ exactly.   

Inserting the leading order into the rotation matrix gives
\be
{\bf R}_{LO}(\sigma)=\frac{\alpha\chi}{2\pi b_0}(\hat{\bf k}\hat{\bf E}-\hat{\bf E}\hat{\bf k})
=\frac{\alpha a_0 F(\sigma)}{2\pi}(\hat{\bf k}\hat{\bf E}-\hat{\bf E}\hat{\bf k}) \;,
\ee
where $F$ is given in~\eqref{chiF}. There are two important things to note here. First, we can write $\alpha a_0=\frac{3\rho}{2\chi_0}$,
so, since we consider $\rho=\mathcal{O}(1)$, the rotation matrix is proportional to a large factor for $\chi\ll1$. This means the spin rotates on a much faster time scale compared to the change in the momentum. That could be problematic for numerical computations if we work with ${\bf M}$. However, the second thing to note is that, unlike ${\bf R}$, ${\bf R}_{LO}$ does not depend on $b_0$. And, since ${\bf R}-{\bf R}_{LO}$ is small for $\chi\ll1$, we can factor out all the fast oscillations by only exponentiating ${\bf R}_{LO}$. This works for both the moments $\tilde{\bf M}(b_0,m)$ and the cumulative distribution function of the spectrum ${\bf M}(b_0,x)$. 
Thus, we substitute
\be
{\bf M}=\exp\left\{\int_\sigma^\infty\ud\sigma'{\bf R}_{LO}(\sigma')\right\}\bar{\bf M} 
\ee 
into~\eqref{integroDiffx} and find 
\be\label{EOMbar}
\begin{split}
\frac{\partial\bar{\bf M}}{\partial\sigma}=&-\Delta{\bf R}\cdot\bar{\bf M}-\int_0^1\ud q\bigg\{{\bf M}_L'\cdot\bar{\bf M}(b_0,x)\\
&+\theta(x-q){\bf M}_C\cdot\bar{\bf M}\left([1-q]b_0,\frac{x-q}{1-q}\right)\bigg\} \;,
\end{split}
\ee
where ${\bf M}_L'$ is ${\bf M}_L$ without the $\hat{\bf k}\hat{\bf E}-\hat{\bf E}\hat{\bf k}$ term and
\be
\Delta{\bf R}={\bf R}-{\bf R}_{LO} \;.
\ee
${\bf R}_{LO}$ does not appear in~\eqref{EOMbar}, so $\bar{\bf M}$ does not have fast oscillations. We have, in other words, factored out most of the spin precession. In PIC codes~\cite{Li:2018fcz} this is instead included by solving the Thomas-Bargmann-Michel-Telegdi equation~\cite{ThomasSpin,Bargmann:1959gz}.  

For a linearly polarized field we can write
\be
\int_{-\infty}^\infty\ud\sigma{\bf R}_{LO}=\varphi(\hat{\bf k}\hat{\bf E}_0-\hat{\bf E}_0\hat{\bf k}) \;,
\ee
where
\be\label{varphiDef}
\varphi=\frac{\alpha a_0}{2\pi}\int_{-\infty}^\infty\ud\sigma\epsilon(\sigma)F(\sigma) \;.
\ee
Since $(\hat{\bf k}\hat{\bf E}_0-\hat{\bf E}_0\hat{\bf k})^2=-(\hat{\bf E}_0\hat{\bf E}_0+\hat{\bf k}\hat{\bf k})$, we have
\be\label{rotCosSin}
\begin{split}
&\exp\left\{\int_\sigma^\infty\ud\sigma'{\bf R}_{LO}(\sigma')\right\}={\bf e}_0{\bf e}_0+\hat{\bf B}_0\hat{\bf B}_0\\
&+(\hat{\bf E}_0\hat{\bf E}_0+\hat{\bf k}\hat{\bf k})\cos\varphi+(\hat{\bf k}\hat{\bf E}_0-\hat{\bf E}_0\hat{\bf k})\sin\varphi \;.
\end{split}
\ee
For $\rho=\mathcal{O}(1)$ we have $\varphi=\mathcal{O}(1/\chi)$, and then even quite small changes in the pulse shape can lead to $\mathcal{O}(1)$ changes in~$\varphi$, which, although relatively small compared to $\varphi=\mathcal{O}(1/\chi)$, can completely change the values of $\cos\varphi$ and $\sin\varphi$. For an oscillating field, the integral in~\eqref{varphiDef} can be identically zero. However, even a small deviation from such a symmetrically oscillating field could lead to $\mathcal{O}(1)$ changes to $\varphi$.

\subsection{Degree of polarization from Frobenius norm}

If the initial Stokes vector is given by
\be
{\bf N}_0={\bf e}_0+p_B\hat{\bf B}_0+p_{Ek}{\bf n} \;,
\ee
where $p_B$ and $p_{Ek}$ are degrees of polarization and 
\be
{\bf n}(\mu)=\cos(\mu)\hat{\bf E}_0+\sin(\mu)\hat{\bf k} 
\ee
is a unit vector in the spin-rotation plane, then from~\eqref{rotCosSin} we find upon projecting the Mueller matrix as in~\eqref{N0MNf}
\be
{\bf n}(\mu)\cdot\exp\left\{\int_\sigma^\infty\ud\sigma'{\bf R}_{LO}(\sigma')\right\}={\bf n}(\mu-\varphi) \;.
\ee
Thus, the effect of ${\bf R}_{LO}$ is to rotate the initial Stokes vector in the $\hat{\bf E}$-$\hat{\bf k}$ plane. Each component of ${\bf n}(\mu-\varphi)\cdot\bar{\bf M}$ will be proportional to either $\cos(\mu-\varphi)$ or $\sin(\mu-\varphi)$ and will therefore be sensitive to the value of $\varphi$. It can therefore be difficult to predict in which direction in the $\hat{\bf E}$-$\hat{\bf k}$ plane the final spin will be, and if one makes an average over experimental runs with slightly different conditions then these components will tend to average out. However, for the degree of polarization we have
\be
\begin{split}
p_f^2:=&[{\bf n}(\mu-\varphi)\cdot\bar{\bf M}\cdot{\bf n}(\nu)]^2\\
&+[{\bf n}(\mu-\varphi)\cdot\bar{\bf M}\cdot{\bf n}(\nu+\pi/2)]^2\\
=&[M_{EE}^2+M_{Ek}^2]\cos^2(\mu-\varphi)\\
&+[M_{kE}^2+M_{kk}^2]\sin^2(\mu-\varphi)\\
&+[M_{EE}M_{kE}+M_{Ek}M_{kk}]\sin[2(\mu-\varphi)] \;,
\end{split}
\ee  
which averages to
\be
\begin{split}
\langle p_f^2\rangle=&\frac{1}{2}[M_{EE}^2+M_{Ek}^2+M_{kE}^2+M_{kk}^2]\\
&=\frac{1}{2}\text{tr}(\bar{\bf M}_{(\LCperp)}^T\bar{\bf M}_{(\LCperp)})=\frac{1}{2}||\bar{\bf M}_{(\LCperp)}||_F^2 \;,
\end{split}
\ee
where $\bar{\bf M}_{(\LCperp)}$ is the part of $\bar{\bf M}$ in the $\hat{\bf E}$-$\hat{\bf k}$ plane and $||...||_F$ is the Frobenius norm. Thus, the degree of polarization is not sensitive to $\varphi$.

We write 
\be
\Delta{\bf R}=(\hat{\bf k}\hat{\bf E}_0-\hat{\bf E}_0\hat{\bf k})\Delta R \;.
\ee
To leading order we have
\be
\Delta R=\frac{\alpha}{b_0}\epsilon(\sigma)\frac{\chi^3}{2\pi}\left\{12\left(\ln\left[\frac{\chi}{\sqrt{3}}\right]-\gamma_E\right)+37\right\} \;.
\ee
Since
\be
\frac{\alpha}{b_0}\epsilon\chi^3=\frac{3}{2}\chi_0\rho\epsilon(\sigma)F^3(\sigma)
\ee
we have to leading order in~\eqref{EOMbar}
\be\label{DeltaRW}
-\Delta{\bf R}\cdot\bar{\bf W}\approx-\frac{\Delta{\bf R}}{(1+\rho J(\sigma))^m} \;.
\ee
We can now calculate $\bar{\bf W}$ to leading order in $\chi\ll1$ in the same way as in Sec.~\ref{LCF of moments}. \eqref{DeltaRW} gives off-diagonal elements that are on the same order of magnitude as the other $\mathcal{O}(\chi)$ terms (we do not consider large $\ln\chi$, because then $\chi$ would have to be extremely small). However, we can still neglect these off-diagonal terms if we only consider the degree of polarization, because of the following. Consider for simplicity $m=0$, then (again with $\rho=\mathcal{O}(\chi_0^0)$)
\be\label{barW12}
\bar{\bf W}_{(\LCperp)}={\bf1}+\chi_0\bar{\bf W}_{(\LCperp)}^{(1)}+\chi_0^2\bar{\bf W}_{(\LCperp)}^{(2)}+\mathcal{O}(\chi_0^3)
\ee
and
\be
\langle p_f^2\rangle=\frac{1}{2}||\bar{\bf M}_{(\LCperp)}||_F^2=1+\chi_0\text{tr}\bar{\bf W}_{(\LCperp)}^{(1)}+\mathcal{O}(\chi_0^2) \;,
\ee
so only the diagonal components contribute to leading order. We find
\be
\begin{split}
\bar{\bf W}_{(\LCperp)}^{(1)}=&
-
\frac{3}{2}\frac{5\sqrt{3}}{8}\int_{-\infty}^\infty\ud\sigma\frac{\rho F^3(\sigma)}{[1+\rho I(\sigma)]^2}
\left(\hat{\bf E}_0\hat{\bf E}_0+\frac{7}{9}\hat{\bf k}\hat{\bf k}\right) \\
&+\dots \hat{\bf E}_0\hat{\bf k}+\dots\hat{\bf k}\hat{\bf E}_0
\;,
\end{split}
\ee
and taking the trace gives
\be
\langle p_f^2\rangle=1-\frac{5\chi_0}{\sqrt{3}}\int_{-\infty}^\infty\ud\sigma\frac{\rho F^3(\sigma)}{[1+\rho I(\sigma)]^2} \;.
\ee
Comparing this with~\eqref{w01}, we see that the degree of depolarization is the same order of magnitude along $\hat{\bf B}_0$ and in the $\hat{\bf E}_0$-$\hat{\bf k}$ plane.

\begin{figure*}
\includegraphics[width=0.49\linewidth]{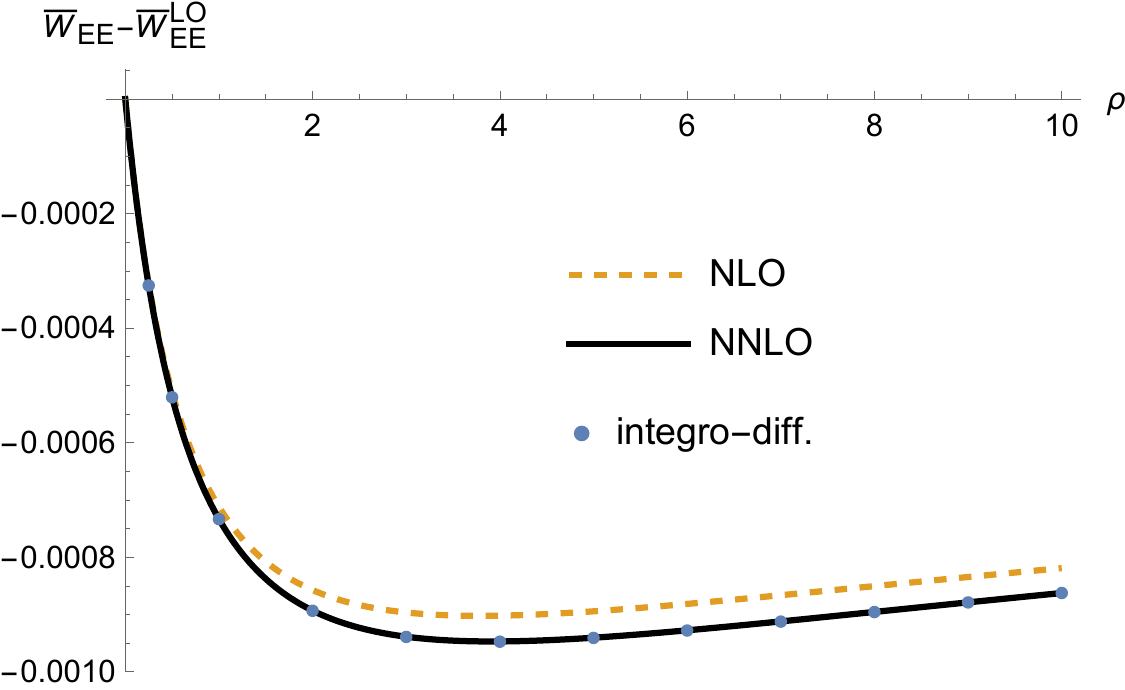}
\includegraphics[width=0.49\linewidth]{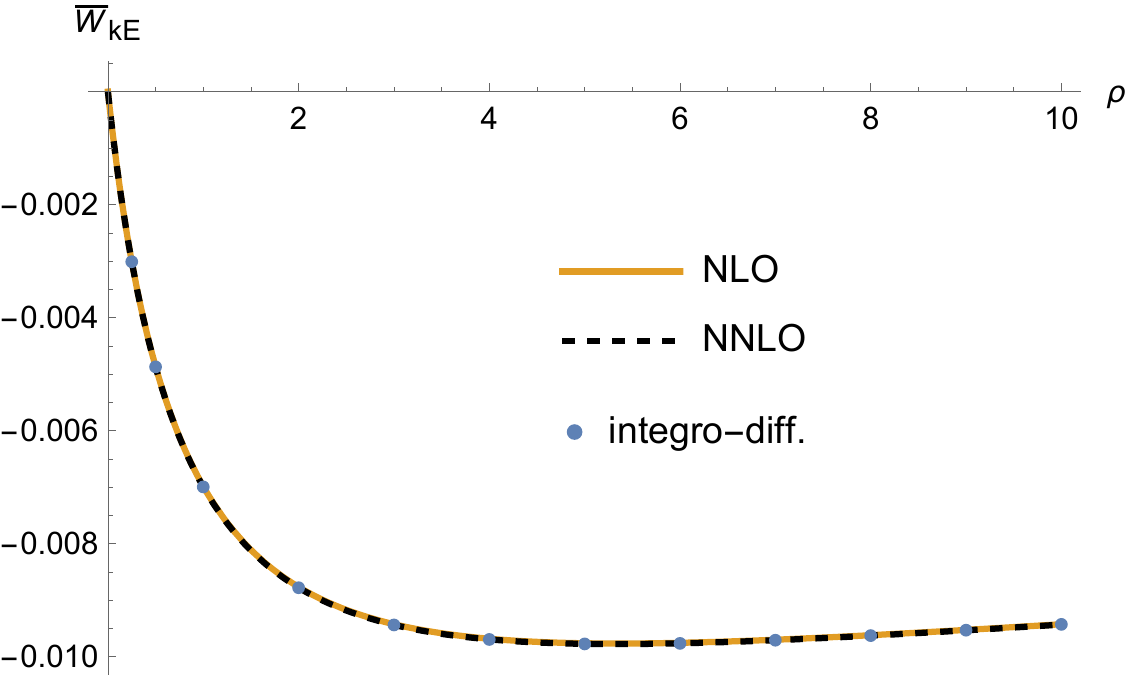}
\caption{$\hat{\bf E}\cdot\bar{\bf W}\cdot\hat{\bf E}$ and $\hat{\bf k}\cdot\bar{\bf W}\cdot\hat{\bf E}$ for a Sauter pulse, $\chi_0=10^{-3}$ and $m=0$. NLO is obtained from~\eqref{baromega1}, and NNLO includes also~\eqref{baromega2}.}
\label{barWfig}
\end{figure*}

\subsection{NLO and NNLO}

In this section we will present the results for the first two orders in~\eqref{barW12} for general $m$, assuming for simplicity a non-oscillating field. In this subsection we will only consider the components in the $\hat{\bf E}-\hat{\bf k}$ plane, so we omit $\perp$ for simplicity. We order to components so that $\hat{\bf E}\to(1,0)$ and $\hat{\bf k}\to(0,1)$ within this 2D space.
We use notation similar to Sec.~\ref{Higher moments},
\be\label{barWsumw}
\bar{\bf W}(\sigma,\chi_0,m)=\sum_{k=0}^\infty\bar{\bf w}_{m,k}(\sigma,\rho,\ln\chi_0)\chi_0^k \;,
\ee
where
\be
\bar{\bf w}_{m,k}=\frac{1}{\rho^{m+k}}\bar{\bm\omega}_{m,k} 
\qquad
\bar{\bm\omega}_{m,0}=\frac{\bf1}{y^m} \;.
\ee
We find (omitting the subscript $m$, $\bar{\bm\omega}_{m,k}\to \bar{\bm\omega}_k$, which is the same on all factors of $\bar{\bm\omega}$)
\begin{widetext}
\be\label{baromega1}
\begin{split}
\bar{\bm\omega}_1=&\int_\sigma^\infty\ud\varphi\, F^3(\varphi)\bigg\{
\frac{55}{32\sqrt{3}}\partial_y^2\bar{\bm\omega}_0-\frac{55}{8\sqrt{3}}\frac{\partial_y\bar{\bm\omega}_0}{y-J(\varphi)}
-\frac{15\sqrt{3}}{16}\begin{pmatrix}1&0\\0&\frac{7}{9}\end{pmatrix}\cdot\frac{\bar{\bm\omega}_0}{[y-J(\varphi)]^2}\\
&\hspace{2.5cm}+\frac{9}{\pi}\left(L+\ln\left[\frac{F(\varphi)}{\sqrt{3}[y-J(\varphi)]}\right]-\gamma_E+\frac{37}{12}\right)\begin{pmatrix}0&-1\\1&0\end{pmatrix}\cdot\frac{\bar{\bm\omega}_0}{[y-J(\varphi)]^2}
\bigg\} \;,
\end{split}
\ee
\be\label{baromega2}
\begin{split}
\bar{\bm\omega}_2=&\int_\sigma^\infty\ud\varphi\, F^3(\varphi)\bigg\{
\frac{55}{32\sqrt{3}}\partial_y^2\bar{\bm\omega}_1-\frac{55}{8\sqrt{3}}\frac{\partial_y\bar{\bm\omega}_1}{y-J(\varphi)}
-\frac{15\sqrt{3}}{16}\begin{pmatrix}1&0\\0&\frac{7}{9}\end{pmatrix}\cdot\frac{\bar{\bm\omega}_1}{[y-J(\varphi)]^2}\\
&\hspace{2.5cm}+\frac{9}{\pi}\left(L+\ln\left[\frac{F(\varphi)}{\sqrt{3}[y-J(\varphi)]}\right]-\gamma_E+\frac{37}{12}\right)\begin{pmatrix}0&-1\\1&0\end{pmatrix}\cdot\frac{\bar{\bm\omega}_1}{[y-J(\varphi)]^2}
\bigg\}\\
+&\int_\sigma^\infty\ud\varphi\, F^4(\varphi)\bigg\{\frac{7}{6}\partial_y^3\bar{\bm\omega}_0-7\frac{\partial_y^2\bar{\bm\omega}_0}{y-J}+
\begin{pmatrix}21&0\\0&22\end{pmatrix}\cdot\frac{\partial_y\bar{\bm\omega}_0}{[y-J]^2}+
\begin{pmatrix}18&0\\0&15\end{pmatrix}\cdot\frac{\bar{\bm\omega}_0}{[y-J]^3}
\bigg\} \;,
\end{split}
\ee
\end{widetext}
where $L=\ln(\chi_0/\rho)$ should be kept constant when differentiating with respect to $y$. We see in Fig.~\ref{barWfig} that for the diagonal element, $\hat{\bf E}\cdot\bar{\bf W}\cdot\bar{\bf E}$, we need to include $\bar{\omega}_2$ to obtain a good precision even for $\chi_0=10^{-3}$, which is quite small. For a Sauter pulse the rotation angle~\eqref{varphiDef} is equal to $\varphi=3\rho/(2\pi\chi_0)$, so $\varphi\sim477\rho$ for $\chi_0=10^{-3}$.

\subsection{Transseries}

From the above we see that, for a linearly polarized field and an electron which has spin components in the $\hat{\bf E}-\hat{\bf k}$ plane, the $\chi_0\ll1$ expansion is not just a power series, it is a transseries with powers $\chi_0^k$, powers of logarithms $[\ln\chi_0]^k$, and terms proportional to either $\cos(\dots/\chi_0)$ or $\sin(\dots/\chi_0)$. For a linearly polarized field, these spin components completely decouple from the ${\bf e}_0-\hat{\bf B}$ components, so if one averages and sums over the initial and final spin then the $\hat{\bf E}-\hat{\bf k}$ components do not contribute and one is left with a $\chi_0\ll1$ expansion which only has powers $\chi_0^k$. However, for a field which is not linearly polarized, the local direction of the magnetic field at one lightfront time is no longer orthogonal to the local direction of the electric field at some other lightfront time, $\hat{\bf B}(\sigma_1)\cdot \hat{\bf E}(\sigma_2)\ne0$, so one can in general expect coupling between the $\hat{\bf E}-\hat{\bf k}$ and ${\bf e}_0-\hat{\bf B}$ components, which means the $[\ln\chi_0]^k$, $\cos(\dots/\chi_0)$ and $\sin(\dots/\chi_0)$ terms in the $\hat{\bf E}-\hat{\bf k}$ space will induce such terms also in the ${\bf e}_0-\hat{\bf B}$ space. 

Consider for example a train of three Sauter pulses
\be\label{threeSauter}
\begin{split}
{\bf a}(\sigma)=a_0[&{\bf e}_x\tanh(\sigma+\Delta\sigma)+{\bf e}_y\tanh(\sigma)\\
&+{\bf e}_x\tanh(\sigma-\Delta\sigma)] \;,
\end{split}
\ee   
where $\Delta\sigma$ is large enough so that the three pulses are well separated. If we sum over the initial and final spin then we should project, ${\bf N}_0\cdot{\bf W}\cdot{\bf N}_f$ with ${\bf N}_0={\bf N}_f={\bf e}_0$. When we integrate backwards from $\sigma=+\infty$ to $\sigma=\Delta\sigma/2$, only the last term in~\eqref{threeSauter} contributes. In this interval the field is approximately linearly polarized with $\hat{\bf B}(\sigma)\propto{\bf e}_y$, so from~\eqref{omegam1} we see that
\be\label{WlastPulse}
{\bf W}(\Delta\sigma/2)\cdot{\bf e}_0=c_0{\bf e}_0+c_y{\bf e}_y \;,
\ee
where $c_0$ and $c_y$ are two nonzero scalars. Then we use~\eqref{WlastPulse} as ``initial'' condition for integrating through the middle pulse, from $\sigma=+\Delta\sigma/2$ to $\sigma=-\Delta\sigma/2$, where $\hat{\bf B}(\sigma)\propto{\bf e}_x$. The ${\bf e}_0$ term in~\eqref{WlastPulse} will generate a ${\bf e}_0$ and a ${\bf e}_x$ term, and the ${\bf e}_y$ term will oscillate as in~\eqref{rotCosSin} in the ${\bf e}_y-{\bf e}_z$ plane. Hence, all components of 
\be\label{WmiddlePulse}
{\bf W}(-\Delta\sigma/2)\cdot{\bf e}_0=d_0{\bf e}_0+d_x{\bf e}_x+d_y{\bf e}_y+d_z{\bf e}_z
\ee 
are in general nonzero. However, when we use~\eqref{WmiddlePulse} as ``initial'' condition for integrating through the first pulse, from $\sigma=-\Delta\sigma/2$ to $\sigma=-\infty$,
the ${\bf e}_x$ and ${\bf e}_z$ components decouple from the other two and will therefore not contribute to ${\bf e}_0\cdot{\bf W}(-\infty)\cdot{\bf e}_0$. The ${\bf e}_0$ component in~\eqref{WmiddlePulse} does contribute to ${\bf e}_0\cdot{\bf W}(-\infty)\cdot{\bf e}_0$ but does not contain any term with $\cos\varphi$ or $\sin\varphi$. We are therefore interested in
\be
{\bf e}_y\cdot{\bf W}(-\Delta\sigma/2)\cdot{\bf e}_0\approx\cos\varphi\;{\bf e}_y\cdot{\bf W}(\Delta\sigma/2)\cdot{\bf e}_0 \;,
\ee    
which will give ${\bf e}_0\cdot{\bf W}(-\infty)\cdot{\bf e}_0$ a term proportional to $\cos\varphi$. To leading order in $\chi_0\ll1$, we can obtain ${\bf e}_y\cdot{\bf W}(\Delta\sigma/2)\cdot{\bf e}_0$ directly from~\eqref{omegam1}. Then we use~\eqref{omegam1} a second time, but replace ${\bm\omega}_{m,0}={\bf1}/y^m$ with
\be
{\bf e}_y\cdot{\bm\omega}_{m,0}\cdot{\bf e}_y\to{\bf e}_y\cdot{\bf W}(-\Delta\sigma/2)\cdot{\bf e}_0 \;,
\ee 
while ${\bf e}_0\cdot{\bm\omega}_{m,0}\cdot{\bf e}_y=0$ still. In other words, we use~\eqref{omegam1} first for the third pulse and then for the first pulse, but for the first pulse ${\bm\omega}_{m,0}$ is more nontrivial due to the more nontrivial ``initial'' conditions at $\sigma=-\Delta\sigma/2$ compared to those at $\sigma=+\infty$. We have thus found that ${\bf e}_0\cdot{\bf W}(-\infty)\cdot{\bf e}_0$ contains a term proportional to $\chi_0^2\cos\varphi$ where $\varphi\propto1/\chi_0$. Although this is suppressed by a factor of $\chi_0^2$ compared to the zeroth order, i.e. the solution to LL, it has fast oscillations which are absent at $\mathcal{O}(\chi_0^0)$ and $\mathcal{O}(\chi_0)$.

\section{Conclusions}

We have derived analytical approximations for the low-energy expansion of $\alpha$-resummed Mueller matrices, where the LO is given by the solution to the Landau-Lifshitz equation. We have checked that these analytical approximations agree with the full numerical solution of the integrodifferential equations for the moments. We have considered the locally constant, locally monochromatic as well as more general regimes. We have showed how to factor out the fast spin precession to obtain an equation with only slowly varying terms. 

We have also shown how to obtain the spectrum from the moments. This is useful since in our approach it is easier and faster to obtain the moments. We have treated $m$ as a continuous, complex variable, which allowed us to obtain a $\chi\ll1$ expansion of the spectrum by performing an inverse Mellin transform. It would be interesting to go beyond this $\chi\ll1$ expansion and perform the Mellin integral numerically and see how much time that would take compared to solving~\eqref{integroDiffx} directly. We have also shown how to use the principle of maximum entropy to obtain a good approximation of the spectrum with just a handful of integer moments. Although considering integer moments already works well, it would be interesting to see if one could find an even faster convergence by considering non-integer moments or some more general expectation values, $\langle f(kP)\rangle$. This, and finding suitable prior distributions, will probably be useful when applying the principle of maximum entropy to cases where the spectrum looks quite different from a Gaussian, e.g. at early times where there is a very sharp peak near $x=0$ or for large $\chi$.

\acknowledgements

G. T. is supported by the Swedish Research Council, Contract No. 2020-04327. Thanks to Tom Blackburn for correspondence about~\cite{Blackburn:2023dey}.

\appendix

\section{LCF of spectrum}\label{LCF of spectrum}

$x$ is related to the final momentum as in~\eqref{xdef}.
The cumulative function ${\bf M}(b_0,x)$ can be obtained from~\cite{Torgrimsson:2023rqo}
\be\label{integroDiffx}
\begin{split}
\frac{\partial{\bf M}}{\partial\sigma}=&-\int_0^1\ud q\bigg\{{\bf M}^L\!\cdot\!{\bf M}(b_0,x)\\
&+\theta(x-q){\bf M}^C\!\cdot\!{\bf M}\left([1-q]b_0,\frac{x-q}{1-q}\right)\bigg\} \;,
\end{split}
\ee
with
\be
{\bf M}(\sigma\to+\infty)={\bf M}^{(0)}={\bf 1} \;.
\ee
To obtain a $\chi\ll1$ approximation of the spectrum, we start by differentiating~\eqref{integroDiffx} with respect to $x$, to obtain an equation directly in terms of the spectrum ${\bf S}=\partial_x{\bf M}$,
\be\label{generalSeq}
\begin{split}
\frac{\partial{\bf S}}{\partial\sigma}=&-\int_0^1\ud q\bigg\{{\bf M}_L\cdot{\bf S}(b_0,x)\\
&+\frac{\theta(x-q)}{1-q}{\bf M}_C\cdot{\bf S}\left[(1-q)b_0,\frac{x-q}{1-q}\right]\bigg\} \\
&-{\bf M}_C(b_0,q=x)\cdot{\bf M}[(1-x)b_0,0] \;.
\end{split}
\ee
This is the generalization of Eq.~(94) in~\cite{Torgrimsson:2023rqo}, which is the corresponding equation for a constant field, to nonconstant fields. After some initial period of time, ${\bf M}[(1-x)b_0,0]$ becomes exponentially suppressed and we can hence neglect it in~\eqref{generalSeq}. To find an approximation we make a similar ansatz as in~\cite{Torgrimsson:2023rqo},
\be\label{Sansatz}
{\bf S}=\frac{e^{-X^2}}{\sqrt{\pi\chi_0\lambda}}\sum_{n=0}^\infty{\bm\rho}_n\chi_0^{n/2} \;,
\ee 
where, instead of the spectrum variable $x$, we use
\be
X=\frac{x-x_{cl}}{\sqrt{\chi_0\lambda}} \;.
\ee
$X$, $\lambda$ and ${\bm\rho}$ only depend on $\chi_0$ via $\rho$, but in this expansion we consider $\rho=\mathcal{O}(1)$. We can think of $\rho$ as independent of the expansion parameter $\chi_0$, but we have to remember to still make the shift $\rho\to(1-q)\rho$ in the second term in~\eqref{generalSeq}.  
We could determine everything about this expansion of ${\bf S}$ without relying on any previous knowledge about ${\bf S}$, but we do in fact already know some things about ${\bf S}$ which allow us to skip some steps. For example, we could determine $x_{cl}$, but we already know that it has to be consistent with the solution to LL, which means
\be\label{xcl}
x_{cl}=1-\frac{kP_{LL}}{b_0}=\frac{\rho J(\sigma)}{1+\rho J(\sigma)} \;.
\ee
We have also used the fact that this expansion has to reduce to the constant-field result in~\cite{Torgrimsson:2023rqo} to write $\mathcal{E}(X^2)=e^{-X^2}/\sqrt{\pi}$. We also set $\rho_0={\bf 1}$.  
The ultimate justification for~\eqref{Sansatz} is the fact that it works, i.e. if the spectrum did not have this mathematical structure we would not be able to find functions $\lambda$ and ${\bm\rho}$ such that it solves~\eqref{generalSeq} to each order in the expansion in $\chi_0$.  

Expanding~\eqref{generalSeq} we find that the terms which would otherwise have been the leading order vanish since we have already used~\eqref{xcl}. At next order in $\mathcal{O}(\sqrt{\chi_0})$, ${\bm\rho}_0={\bm1}$ is still the only ${\bm\rho}_n$ that contributes at this order, which means there is only one rather than four equations, and things simplify further compared to~\cite{Torgrimsson:2023rqo} since we have already used $\mathcal{E}(X^2)=e^{-X^2}/\sqrt{\pi}$. After dividing the equation by $e^{-X^2}$, we find one term that does not depend on $X$ and one that is quadratic in $X$, $f_0(\sigma,\rho)+f_2(\sigma,\rho)X^2=0$. Both terms have to be zero independently, since $x$ only appears in $X$. But at this order we only have one unknown function, $\lambda$.  Fortunately, we find $f_2(\sigma,\rho)/f_0(\sigma,\rho)=-2$, so if $f_0(\sigma,\rho)=0$ then $f_2(\sigma,\rho)=0$ automatically. This is one consistency check that~\eqref{Sansatz} is correct. We thus have one independent equation for one unknown,
\be
\partial_\sigma\lambda-\rho F^2[\rho\partial_\rho\lambda+3\lambda]+\frac{55}{8\sqrt{3}}(1-x_{cl})^4\rho F^3=0 \;.
\ee     
To make this an ordinary differential equation, we again change variables from $\sigma$ and $\rho$ to $\sigma$ and $y$ in~\eqref{characteristicy},
\be
\partial_\sigma\frac{\lambda(\sigma,y)}{[y-J(\sigma)]^3}=-\frac{55F^3(\sigma)}{8\sqrt{3}y^4} \;.
\ee
Thus, $\lambda$ is given by~\eqref{lambdaLCF}.
We have $\lambda(\sigma\to-\infty)=2S^2/\chi_0$, where the standard deviation $S^2$ is given by~\eqref{standardLCF}.

\subsection{Comparison with literature}\label{Comparison with literature}

As shown in~\cite{Torgrimsson:2023rqo}, once we have obtained the result for a wave packet which is initially sharply (infinitely) peaked, we can use that result as a Green's function for obtaining the corresponding result for cases when the initial wave packet has a finite width. Therefore, to compare with the results in~\cite{Blackburn:2023dey} we set the initial variance to zero.
We immediately see that~\eqref{standardLCF} agrees with Eq.~(14) in~\cite{Blackburn:2023dey}, which was obtained there with very different approach.

The $\{1,0\}\cdot{\bf w}_{1,1}\cdot\{1,0\}$ component in~\eqref{w11} gives the first moment after averaging and summing over initial and final spin. We can compare this component with the equations in~\cite{Blackburn:2023dey} as follows. Eq.~(10) in~\cite{Blackburn:2023dey} gives an ordinary, rather than partial, differential equation for the first moment, denoted there as $\mu$, coupled to higher-order moments, the variance $\sigma$ etc. After expanding that equation to first order in $\chi_0$ one finds an equation which couple $\mu$ to only $\sigma$ and no higher-order moments,
\be\label{muEq}
\begin{split}
0=&-\hat{\mu}_{(0)}'-\frac{2}{3}R_c|f|^2\hat{\mu}_{(0)}^2
+\chi_0\bigg(-\hat{\mu}_{(1)}'\\
&-\frac{2}{3}R_c|f|^2[2\hat{\mu}_{(1)}\hat{\mu}_{(0)}+\hat{\sigma}_{(1)}^2]+\frac{55R_c}{8\sqrt{3}}|f|^3\hat{\mu}_{(0)}^3\bigg) \;,
\end{split}
\ee 
where we have used essentially the same notation as in~\cite{Blackburn:2023dey}, so to translate to our notation use $(2/3)R_c\to\rho$ and $|f|\to F$. The first two terms determine $\hat{\mu}_{(0)}$ as the solution to LL, and $\hat{\sigma}_{(1)}$ is given by Eq.~(14) in~\cite{Blackburn:2023dey}. One can then integrate the $\mathcal{O}(\chi_0)$ part of~\eqref{muEq} to find $\hat{\mu}_{(1)}$. Since $\hat{\sigma}_{(1)}$ already contains an integral over $\phi$, one might have guessed that the result would contain more integrals compared to~\eqref{w11}, but it is possible to simplify using partial integration.
The resulting $\hat{\mu}_{(1)}$ agrees exactly with the $\{1,0\}\cdot{\bf w}_{1,1}\cdot\{1,0\}$ component in~\eqref{w11}.

\end{document}